# Analysis of Kerr comb generation in silicon microresonators under the influence of two-photon absorption and free-carrier absorption

P. Trocha,[1,*] J. Gärtner,[2] P. Marin-Palomo,[1]
W. Freude,[1] W. Reichel,[2] C. Koos,[1,3,**]
[1]*Institute of Photonics and Quantum Electronics (IPQ), Karlsruhe Institute of Technology (KIT), 76131 Karlsruhe, Germany*
[2]*Institute of Analysis (IANA), Karlsruhe Institute of Technology (KIT), 76131 Karlsruhe, Germany*
[3]*Institute of Microstructure Technology (IMT), Karlsruhe Institute of Technology (KIT), 76131 Germany*
*philipp.trocha@kit.edu, **christian.koos@kit.edu

Kerr frequency comb generation relies on dedicated waveguide platforms that are optimized towards ultra-low loss while offering comparatively limited functionality restricted to passive building blocks. In contrast to that, the silicon-photonic platform offers a highly developed portfolio of high-performance devices, but is deemed to be inherently unsuited for Kerr comb generation at near-infrared (NIR) telecommunication wavelengths due to strong two-photon absorption (TPA) and subsequent free-carrier absorption (FCA). Here we present a theoretical investigation that quantifies the impact of TPA and FCA on Kerr comb formation and that is based on a modified version of the Lugiato-Lefever equation (LLE). We find that silicon microresonators may be used for Kerr comb generation in the NIR, provided that the dwell time of the TPA-generated free-carriers in the waveguide core is reduced by a reverse-biased *p-i-n*-junction and that the pump parameters are chosen appropriately. We validate our analytical predictions with time integrations of the LLE, and we present a specific design of a silicon microresonator that may even support formation of dissipative Kerr soliton combs.

## I. INTRODUCTION

Generation of optical frequency combs in high-Q Kerr-nonlinear microresonators [1,2] has the potential to unlock a wide range of applications such as timekeeping [3], frequency synthesis [4], optical communications [5], spectroscopy [6] and optical ranging [7,8]. Amongst different frequency comb states, dissipative Kerr solitons (DKS) [9] are particularly attractive, offering broadband optical spectra with hundreds of phase-locked optical tones spaced by free spectral ranges of tens of gigahertz. As a key advantage in comparison to conventional comb sources built from discrete components, Kerr comb generators offer small footprint and can be integrated into robust chip-scale photonic systems that lend themselves to cost-efficient mass production. So far, integrated optical Kerr comb sources have mostly been realized using specifically optimized silica and silicon-nitride-based waveguides that offer ultra-low propagation loss down to $5.5\,\mathrm{dB\,m^{-1}}$ along with anomalous group-velocity dispersion [10,11], and that can bear high power levels. The functionality of these integration platforms, however, is still limited to merely passive devices. In contrast to that, silicon photonics offers a highly developed portfolio of active and passive devices that are specifically geared towards operation at near-infrared (NIR) telecommunication wavelengths between $1200\,\mathrm{nm}$ and $1700\,\mathrm{nm}$. These devices can be reliably fabricated at low cost on large-area silicon substrates [12–15] and lend themselves to co-integration with electronic devices [16]. Expanding the silicon-photonic integration platform by monolithically integrated Kerr comb sources could have transformative impact regarding functionality, performance, footprint and cost of comb-based optical systems. However, the current understanding is that silicon microresonators are inherently unsuited for Kerr comb generation at NIR telecommunication wavelengths due to strong two-photon absorption (TPA) and subsequent free-carrier absorption (FCA) [17,18]. While simulations for specific parameter sets confirm this notion [17], a broad theoretical investigation of Kerr comb formation under the influence of TPA and FCA is still lacking.

In this paper we present a theoretical analysis of the impact of TPA and FCA on Kerr comb formation. We build upon an analytical model that expands the Lugiato-Lefever equation (LLE [19,20]) to include TPA, relaxation dynamics of TPA-induced free carriers, as well as dispersion anomalies of the ring resonances that arise as a consequence of avoided mode crossings. In a first step, we use the TPA coefficient as well as the lifetime and the absorption cross-section of the associated free carriers as central parameters and formulate simple necessary conditions that must be fulfilled for achieving modulation instability (MI) and subsequent comb formation. We describe the dependence of the MI threshold pump power on TPA and FCA parameters, and we find an upper limit for the TPA coefficient, above which comb formation is impossible even in absence of FCA. The theoretical predictions are independently confirmed by numerical simulations that are based on time-integration of the Lugiato-Lefever equation. This model is general and can be broadly applied to different material platforms. In a second step, we use our model to investigate silicon-photonic microresonators, in which the free-carrier dwell time can be artificially reduced by a reverse-biased *p-i-n*-junction that is built around the respective waveguide core. We find that Kerr comb generation in silicon microresonators can be achieved within technically realistic parameter ranges for free-carrier lifetime and pump power. We further develop and numerically validate a design for a silicon-photonic Kerr comb source that has a free spectral range (FSR) of $100\,\mathrm{GHz}$ and a threshold pump power of $12\,\mathrm{mW}$ and that should even be suitable for DKS comb formation.

## II. MODEL

In our model, we describe the time- and space-dependent electric field $E(t)=\Re\left(\underline{E}(t)\,\mathrm{e}^{\mathrm{i}\omega_\mathrm{P}t}\right)$ in the Kerr-nonlinear microresonator (MR) by a complex slowly varying time- and space-dependent amplitude $\underline{E}$ and a carrier at the (angular) frequency $\omega_\mathrm{P}$ of the optical pump wave. The presence of the strong pump leads to parametric gain for a pair of modes located symmetrically to both sides of the pump frequency $\omega_\mathrm{P}$. If the parametric gain for any of these modes exceeds the resonator losses, the corresponding mode amplitudes are amplified by resonant four-wave mixing, drawing energy from the pump wave. At the same time, the presence of a strong pump wave leads to generation of free carriers (FC) through two-photon absorption (TPA). These carriers accumulate and lead to additional optical losses through free-carrier absorption (FCA). For silicon-photonic microresonators, the dwell time $\tau_\mathrm{eff}$ and the associated concentration $N_\mathrm{c}$ of the free carriers can be influenced by a reverse-biased *p-i-n*-junction [21,22], see Fig. 1. In the following, we consider the evolution of the field amplitude and of the free-carrier density over multiple cavity round-trip times $t_\mathrm{RT}=L/v_\mathrm{g}=f_\mathrm{FSR}^{-1}$, where $f_\mathrm{FSR}$ denotes the free spectral range of the cavity as defined by the perimeter $L$ and by the optical group velocity $v_\mathrm{g}=c/n_\mathrm{g}$ that is obtained at the pump frequency $\omega_\mathrm{P}$ in absence of free carriers. The number of round-trips is denoted by an integer $m$, and we introduce a long time scale ("slow" time variable) $t=m\,t_\mathrm{RT}$, which we consider to be continuous. At the same time, we model the evolution of the complex slowly varying amplitude $\underline{E}$ and of the FC density $N_\mathrm{c}$ within the cavity using a short time scale ("fast" time variable) $\tau=t-z/v_\mathrm{g}$ that is retarded according to the position $z$ inside the cavity. The slowly varying amplitude $\underline{E}(t,\tau)$ inside the resonator is modelled as a superposition of fields with complex envelopes $\underline{E}_{\Omega'}(t)$, oscillating at equidistant angular frequencies which are offset from the pump frequency $\omega_\mathrm{P}=\omega_0$ by $\omega_{\Omega'}=\Omega'\times\left(2\pi\times f_\mathrm{FSR}\right)$ $\left(\Omega'=0,\pm1,\pm2,\pm3,\ldots\right)$ and thus given as

$$\underline{E}(t,\tau)=\sum_{\Omega'}\underline{E}_{\Omega'}(t)\,\mathrm{e}^{\mathrm{i}\omega_{\Omega'}\tau}=\sum_{\Omega'}\underline{E}_{\Omega'}(t)\,\mathrm{e}^{\mathrm{i}2\pi\Omega'\tau/t_R}, \tag{1}$$

The imaginary unit is $\mathrm{i}$. The field and the FC density obey periodic boundary conditions, $\underline{E}(t,\tau)=\underline{E}(t,\tau+t_\mathrm{RT})$, $N_\mathrm{FC}(t,\tau)=N_\mathrm{FC}(t,\tau+t_\mathrm{RT})$. Disregarding temperature effects, self-steepening, higher-order dispersion and higher-order multi-photon absorption, the LLE and the FC equation read [17,18]

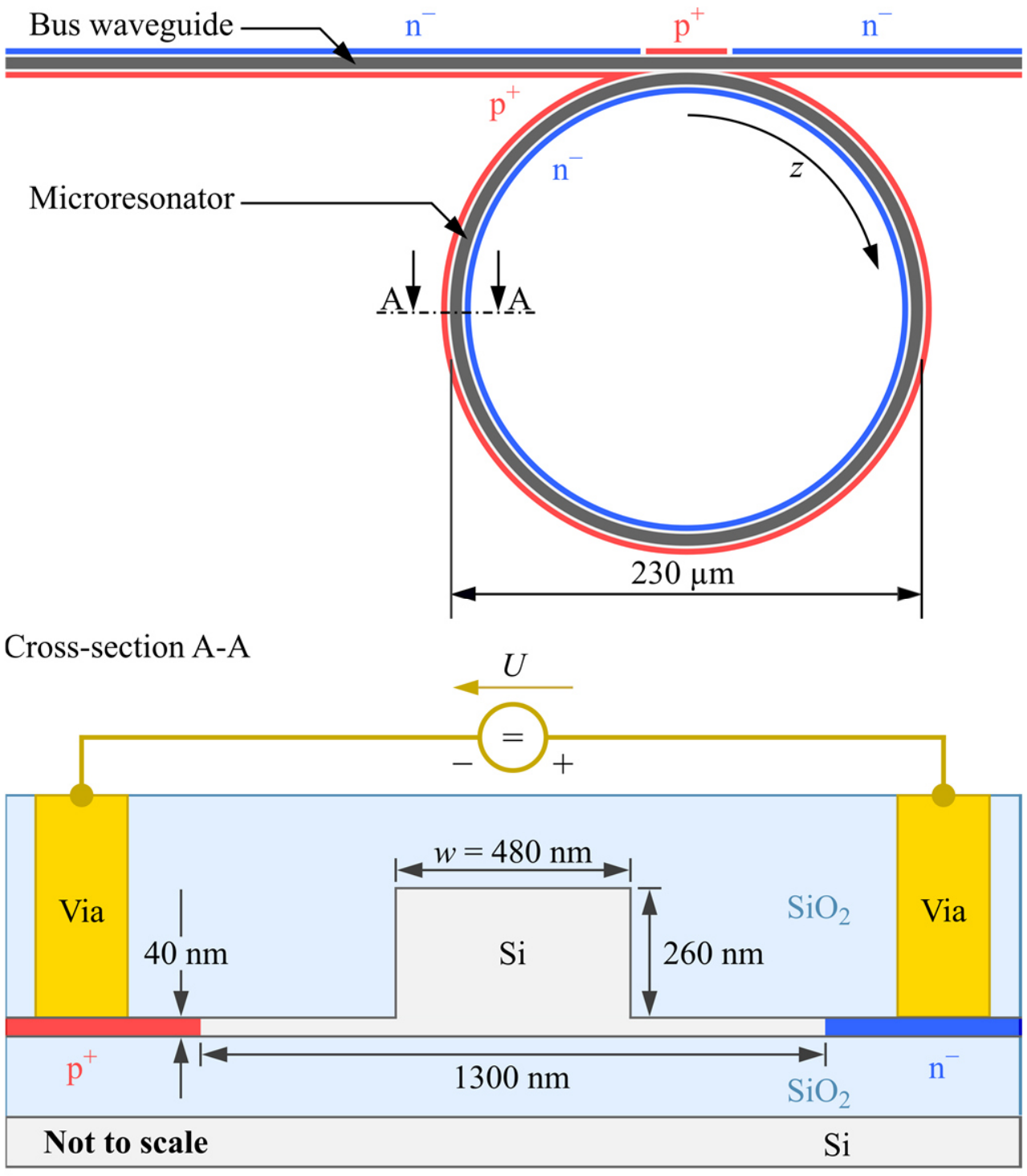

**Figure 1**: Silicon-photonic microresonator as an example of a device suffering from both TPA and FCA. The bus waveguide as well as the resonator ring waveguide are surrounded by $p^{+}$-doped (red) and $n^{-}$-doped regions (blue) that form a *p*-*i*-*n*-junction. This junction allows reducing the dwell time of free carriers by applying a reverse bias voltage $U$ through contact pads and vias. At the coupling section between the bus and the ring waveguide, the doping is locally inverted from an $n^{-}$-doping to a $p^{+}$-doping next to the bus waveguide to ensure maximum free-carrier removal in the microresonator [23]. For the geometrical dimensions indicated in the cross-section, the waveguide features anomalous group-velocity dispersion (GVD) at wavelengths near 1550 nm [24]. In this case a ring diameter of 230 µm leads to a free spectral range of $f_{\mathrm{FSR}} = 100\,\mathrm{GHz}$, corresponding to a round-trip time $t_{\mathrm{RT}} = f_{\mathrm{FSR}}^{-1} = 10\,\mathrm{ps}$.

$$t_{\mathrm{RT}} \frac{\partial \underline{E}(t,\tau)}{\partial t} = \sqrt{\kappa}\underline{E}_{\mathrm{in}} + \left[ -\frac{\alpha_i L}{2} - \frac{\kappa}{2} - \mathrm{i}\delta_0 - \mathrm{i}\frac{1}{2}\beta_2 L \frac{\partial^2}{\partial \tau^2} + \mathrm{i}\hat{\Phi}_{\mathrm{AMC}} + \left( \mathrm{i}\gamma L - \frac{\beta_{\mathrm{TPA}} L}{2A_{\mathrm{eff}}} \right) \left| \underline{E}(t,\tau) \right|^2 - \frac{\sigma \Gamma_{\mathrm{c}} L}{2} (1+\mathrm{i}\mu) N_{\mathrm{FC}}(t,\tau) \right] \underline{E}(t,\tau), \tag{2}$$

$$\frac{\partial N_{\mathrm{c}}(t,\tau)}{\partial \tau} = \frac{\beta_{\mathrm{TPA}}}{2\hbar\omega_{\mathrm{P}}} \frac{\left| \underline{E}(t,\tau) \right|^4}{A_{\mathrm{eff}}^2} - \frac{N_{\mathrm{c}}(t,\tau)}{\tau_{\mathrm{eff}}}. \tag{3}$$

In these relations, the quantity $\underline{E}_{\mathrm{in}}$ denotes the electric field amplitude of the pump with power $P_{\mathrm{in}} = |\underline{E}_{\mathrm{in}}|^2$ in the bus waveguide, see Fig. 1, $\kappa$ describes the coupling to the microresonator, and $\alpha_i$ (unit $\mathrm{m}^{-1}$) is the waveguide power loss coefficient inside the MR. The resonator parameters are assumed to be the same for all complex envelopes $\underline{E}_{\Omega'}(t)$, unless specified otherwise. The detuning $\delta_0 = (\omega_{\mathrm{R}} - \omega_{\mathrm{P}}) t_{\mathrm{RT}}$ corresponds to the offset of the pump frequency $\omega_{\mathrm{P}}$ from the resonance frequency $\omega_{\mathrm{R}}$ of the pumped mode. The coefficient $\beta_2$ describes the second-order dispersion of the cavity. The nonlinearity parameter of the resonator waveguide is denoted by $\gamma = \omega_{\mathrm{P}} c^{-1} n_2 A_{\mathrm{eff}}^{-1}$ (unit $\mathrm{m}^{-1}\,\mathrm{W}^{-1}$) with the nonlinear Kerr coefficient $n_2$ (unit $\mathrm{m}^2\,\mathrm{W}^{-1}$), the vacuum speed of light $c$, and the effective mode-field area $A_{\mathrm{eff}}$. The quantity $\beta_{\mathrm{TPA}}$ denotes the two-photon absorption coefficient, $\sigma$ is the free carrier absorption cross-section, and $\mu$ the free-carrier dispersion parameter which describes the influence of FC on the real part of the refractive index. The reduced Planck constant is $\hbar$. The model additionally includes a field confinement factor $\Gamma_c$ which takes account of the fact that only a fraction of the optical mode field experiences the attenuation by FC generated in the resonator waveguide [25]. Since we consider only modes for which the field is strongly confined to the waveguide, we may assume $\Gamma_c \approx 1$ , whereas other waveguide designs, e. g., slot waveguides with a nonlinear organic cladding [26,27] may lead values of $\Gamma_c$ that are significantly smaller than 1. Finally, we include the possibility of local resonance shifts $\delta\omega_{\Omega'}$ caused by avoided mode crossings (AMC) [28–30]. These resonance shifts lead to additional phase shifts $\delta\omega_{\Omega'} t_{\mathrm{RT}}$ for the respective electric field envelopes $\underline{E}_{\Omega'}(t)$ . The impact of AMC can hence be described by an operator $\hat{\Phi}_{\mathrm{AMC}}$ acting on the envelope field $\underline{E}(t,\tau)$ of Eq. (1),

$$\begin{aligned}\hat{\Phi}_{\mathrm{AMC}}\,\underline{E}(t,\tau) &= \sum\nolimits_{\tilde{\Omega}}\left(\delta\omega_{\tilde{\Omega}}\exp\left(\mathrm{i}\,2\pi\tilde{\Omega}\tau/t_{\mathrm{RT}}\right)\int_0^{t_{\mathrm{RT}}}\underline{E}(t,\tau_1)\exp\left(-\mathrm{i}\,2\pi\tilde{\Omega}\tau_1/t_{\mathrm{RT}}\right)\mathrm{d}\tau_1\right)\\ &= \sum\nolimits_{\tilde{\Omega},\Omega'}\left(\delta\omega_{\tilde{\Omega}}t_{\mathrm{RT}}\underline{E}_{\Omega'}(t)\exp\left(\mathrm{i}\,2\pi\tilde{\Omega}\tau/t_{\mathrm{RT}}\right)\delta_{\tilde{\Omega},\Omega'}\right)\\ &= \sum\nolimits_{\Omega'}\left(\delta\omega_{\Omega'}t_{\mathrm{RT}}\underline{E}_{\Omega'}(t)\exp\left(\mathrm{i}\,2\pi\Omega'\tau/t_{\mathrm{RT}}\right)\right).\end{aligned} \tag{4}$$

where $\delta_{\tilde{\Omega},\Omega'}$ denotes the Kronecker symbol. In absence of AMC, all resonance shifts $\delta\omega_{\Omega'}$ are zero and thus do not lead to any additional phase shifts $\delta\omega_{\Omega'} t_{\mathrm{RT}}$ for the electric field envelopes $\underline{E}_{\Omega'}(t)$ .

In the following, we simplify Eq. (2) by assuming critical coupling , i.e., $\alpha_i L = \kappa$ . For normalization, we multiply Eq. (2) with $\sqrt{\gamma n_{\mathrm{g}}^2/\left(\alpha_i^3 t_{\mathrm{RT}}^2 c^2\right)}$ and Eq. (3) with $n_2\hbar\omega_{\mathrm{P}}^2/\left(\alpha_i^2 c\right)$. We introduce the normalized quantities specified in Table 1 and reformulate Eq. (2), (3) and (4):

$$\frac{\partial\underline{a}(t',\tau')}{\partial t'} = \sqrt{F} + \left[-1-\mathrm{i}\zeta+\mathrm{i}\,d\frac{\partial^2}{\partial\tau'^2}+\mathrm{i}\hat{\Phi}'_{\mathrm{AMC}}+(\mathrm{i}-r)\left|\underline{a}(t',\tau')\right|^2-\sigma'(1+\mathrm{i}\,\mu)N'_{\mathrm{c}}(t',\tau')\right]\underline{a}(t',\tau'), \tag{5}$$

$$\frac{\partial N'_{\mathrm{c}}(t',\tau')}{\partial\tau'} = r\left|\underline{a}(t',\tau')\right|^4 - \frac{N'_{\mathrm{c}}(t',\tau')}{\tau'_{\mathrm{eff}}}, \tag{6}$$

$$\hat{\Phi}'_{\mathrm{AMC}}\underline{a}(t',\tau') = \sum\nolimits_{\tilde{\Omega}}\frac{\phi_{\tilde{\Omega}}}{2\pi}\int_0^{2\pi}\underline{a}(t',\tau_1)\exp\left[-\mathrm{i}\tilde{\Omega}\tau_1\right]\mathrm{d}\tau_1\exp\left[\mathrm{i}\tilde{\Omega}\tau'\right]. \tag{7}$$

To investigate under which circumstances modulation instability can occur when pumping the resonator, we need to know whether any pair of resonator modes experiences a sufficiently high parametric gain to overcome the total resonator loss. This loss includes linear propagation loss and coupling loss, which are equal for critical coupling, $\alpha_i L = \kappa$, and which are expressed by $-1$ in Eq. (5), two-photon absorption, expressed by $-r\left|\underline{a}(t',\tau')\right|^2$ in Eq. (5), as well as free-carrier absorption, expressed by $-\sigma' N'_{\mathrm{c}}(t',\tau')$ in Eq. (5), where the normalized density of accumulated carriers strongly depends on the carrier dwell time $\tau_{\mathrm{eff}}$ and its normalized counterpart $\tau'_{\mathrm{eff}}$, see Eq. (6). For exploring modulation instability, we use an ansatz for the normalized optical resonator field $\underline{a}(t',\tau')$, consisting of a strong field $\underline{a}_0$ in the pumped resonator mode (normalized power $A = \left|\underline{a}_0\right|^2$ ) and a pair of weak fields $\underline{a}_{\pm\Omega}$ (“sidebands” for short) in resonator modes which are offset from the pump frequency $\omega_{\mathrm{P}}$ by $\omega_{\pm\Omega} = \pm\Omega\times(2\pi\times\mathrm{FSR})$ ( $\Omega\in\mathbb{N}_+$ ) [31]. The amplitude of these sidebands may change with time with a complex normalized gain rate $\underline{\lambda} = \lambda + \mathrm{i}\lambda_i$ , $\lambda,\lambda_i\in\mathbb{R}$ , leading to a three-wave ansatz of the form

Table 1: Normalized parameters and physical quantities.

| | | | |
|---|---|---|---|
| Slow time | $t' = \alpha_i L t / t_{\mathrm{RT}} = t\,\alpha_i c / n_{\mathrm{g}}$ | Phase shift operator | $\hat{\Phi}'_{\mathrm{AMC}} = \hat{\Phi}_{\mathrm{AMC}}\, n_{\mathrm{g}} / \left(\alpha_i t_{\mathrm{RT}} c\right)$ |
| Fast time | $\tau' = 2\pi\,\tau / t_{\mathrm{RT}}$ | Phase shifts | $\phi_{\Omega'} = \delta\omega_{\Omega'}\, n_{\mathrm{g}} / \left(\alpha_i c\right)$ |
| Optical field | $\underline{a}(t', \tau') = \sqrt{\gamma/\alpha_i}\,\underline{E}(t, \tau)$ | Dispersion | $d = -2\beta_2 \pi^2 / \left(\alpha_i t_{\mathrm{RT}}^2\right)$ |
| Free carrier density | $N'_{\mathrm{c}}(t', \tau') = 2\pi\hbar\omega^2\, n_2 / \left(c\,\alpha_i^2 t_{\mathrm{RT}}\right) N_{\mathrm{c}}(t, \tau)$ | TPA parameter | $r = c\beta_{\mathrm{TPA}} / \left(2\omega_{\mathrm{P}} n_2\right)$ |
| Pump field | $\sqrt{F} = \sqrt{\gamma P_{\mathrm{in}} n_{\mathrm{g}} / \left(c\,\alpha_i^2 t_{\mathrm{RT}}\right)}$ | FC dwell time | $\tau'_{\mathrm{eff}} = 2\pi\tau_{\mathrm{eff}} / t_{\mathrm{RT}}$ |
| Detuning | $\zeta = \delta_0 n_{\mathrm{g}} / \left(\alpha_i t_{\mathrm{RT}} c\right)$ | FC cross-section | $\sigma' = \alpha_i t_{\mathrm{RT}} \sigma \Gamma_{\mathrm{c}} c / \left(4\pi\hbar\omega_{\mathrm{P}}^2 n_2\right)$ |

$$\underline{a}(t', \tau') = \underline{a}_0 + \underline{\hat{a}}_{+\Omega} \mathrm{e}^{\lambda t'} \mathrm{e}^{\mathrm{i}\lambda_i t'} \mathrm{e}^{\mathrm{i}\Omega\tau'} + \underline{\hat{a}}_{-\Omega} \mathrm{e}^{\lambda t'} \mathrm{e}^{-\mathrm{i}\lambda_i t'} \mathrm{e}^{-\mathrm{i}\Omega\tau'}, \qquad \left|\underline{\hat{a}}_{\pm\Omega}\right| << \left|\underline{a}_0\right|. \tag{8}$$

We assume that the sideband amplitudes $\underline{\hat{a}}_{\pm\Omega}$ are initially much smaller than the amplitude of the pumped mode $\underline{a}_0$, $\left|\underline{\hat{a}}_{\pm\Omega}\right| << \left|\underline{a}_0\right|$, such that we can treat them as a weak perturbation by linearizing Eqs. (5) and (6) in $\underline{\hat{a}}_{+\Omega}$, $\underline{\hat{a}}_{-\Omega}$ about the strong field of the pumped mode $\underline{a}_0$. Inserting Eq. (8) into the linearized version of Eqs. (5) and (6) allows us to derive an expression for the gain rate $\underline{\lambda}(\Omega)$, see Appendix A. Modulation instability occurs for $\lambda(\Omega) > 0$, and the sideband amplitudes at $\pm\Omega$ grow exponentially with time. The field oscillates with $\pm\Omega$ with respect to the normalized fast time $\tau'$, and it experiences a phase shift $\pm\lambda_i t'$ with normalized slow time $t'$.

To identify resonator and pump parameters for which modulation instability can occur, we first derive an expression for $\underline{\lambda}$ in terms of these parameters. To this end, we first solve Eq. (6) for $N'_{\mathrm{c}}(t', \tau')$ and substitute the result in Eq. (5). Next we substitute Eq. (8) in Eq. (5), neglecting the small second-order products of the form $\underline{\hat{a}}_{\pm\Omega}^2, \left|\underline{\hat{a}}_{\pm\Omega}\right|^2, \underline{\hat{a}}_{+\Omega}\underline{\hat{a}}_{-\Omega}$. We solve the resulting equation for $\underline{\lambda}$, see Appendix A, and obtain two complex solutions $\underline{\lambda}_{\pm} = \lambda_{\pm} + \mathrm{i}\,\lambda_{\pm,i}$. From these solutions, we select the one for which the real part $\lambda_+$ can assume positive values, corresponding to modulation instability,

$$\begin{aligned}
\lambda(\Omega) &= -1 - 2rA - \frac{r\tau'_{\mathrm{eff}}\sigma'\left(3 + \left(\Omega\tau'_{\mathrm{eff}}\right)^2\right)}{1 + \left(\Omega\tau'_{\mathrm{eff}}\right)^2} A^2 + \Re\{\underline{\Delta}\} \quad \text{with } \Re\{\underline{\Delta}\} > 0, \\
\lambda_i(\Omega) &= \frac{\phi_{+\Omega} - \phi_{-\Omega}}{2} + 2\frac{r\tau'_{\mathrm{eff}}\sigma'\Omega\tau'_{\mathrm{eff}}}{1 + \left(\Omega\tau'_{\mathrm{eff}}\right)^2} A^2 + \Im\{\underline{\Delta}\}, \\
\underline{\Delta} &= \left[ A^2 \left( r + \frac{2r\tau'_{\mathrm{eff}}\sigma'}{1 + \left(\Omega\tau'_{\mathrm{eff}}\right)^2} A - \mathrm{i}\frac{2r\tau'_{\mathrm{eff}}\sigma'\Omega\tau'_{\mathrm{eff}}}{1 + \left(\Omega\tau'_{\mathrm{eff}}\right)^2} A \right)^2 \right. \\
&\quad + A^2 \left( 1 - \frac{2r\tau'_{\mathrm{eff}}\sigma'\mu}{1 + \left(\Omega\tau'_{\mathrm{eff}}\right)^2} A + \mathrm{i}\frac{2r\tau'_{\mathrm{eff}}\sigma'\Omega\tau'_{\mathrm{eff}}\mu}{1 + \left(\Omega\tau'_{\mathrm{eff}}\right)^2} A \right)^2 \\
&\quad \left. - \left( \zeta + d\Omega^2 - \frac{\phi_{+\Omega} + \phi_{-\Omega}}{2} - 2A + \frac{r\tau'_{\mathrm{eff}}\sigma'\left(3 + \left(\Omega\tau'_{\mathrm{eff}}\right)^2\right)\mu}{1 + \left(\Omega\tau'_{\mathrm{eff}}\right)^2} A^2 - \mathrm{i}\frac{2r\tau'_{\mathrm{eff}}\sigma'\Omega\tau'_{\mathrm{eff}}\mu}{1 + \left(\Omega\tau'_{\mathrm{eff}}\right)^2} A^2 \right)^2 \right]^{1/2}.
\end{aligned} \tag{9}$$

Both the real part and the imaginary part of the gain parameter show a dependence on the normalized power $A$ of the pumped mode. For a given normalized pump power $F$, the normalized power $A$ in the pumped mode can be determined by evaluating the expression

$$F = \left[ \left(1 + rA + r\tau'_{\text{eff}}\,\sigma' A^2\right)^2 + \underbrace{\left(A - \zeta + \phi_0 - r\tau'_{\text{eff}}\,\sigma'\mu A^2\right)^2}_{(*)} \right] A. \tag{10}$$

A derivation of Eq. (10) can be found in Appendix A.

## III. PARAMETER RANGES OF TPA, FCA, AND PUMP POWER LEADING TO MODULATION INSTABILITY

Using Equations (9) and (10) for a specific parameter set $\{r, \sigma', \tau'_{\text{eff}}, \mu, d, \phi_{\tilde{\Omega}}\}$ and for specific operating conditions $\{F, \zeta\}$, we can determine whether MI can occur and at which sidebands $\Omega' = \pm\Omega$ it will happen. To reduce the complexity of the evaluation, we simplify Eq. (9) by considering technically relevant sets of normalized parameters $\{r \approx 1, \sigma' \approx 0.01, \tau'_{\text{eff}} \approx 2\pi \ldots 20\pi, \mu \approx 10, F \approx 10\}$ and by assuming that the side modes, for which MI will occur, are not affected by AMC, i.e., $\phi_{+\Omega} = \phi_{-\Omega} = 0$. The normalized parameters are obtained using the relations in Table 1 in combination with the physical parameters listed in Tables G1 and G2 Appendix G. Specifically, we assume that TPA-generated free carriers are removed by a reversed-biased *p-i-n*-junction and thus have a small dwell time $\tau_{\text{eff}} \approx (12\ldots100)\,\text{ps}$ [21]. For estimating the normalized power $A$ of the pumped mode, we use Eq. (10) and assume that a detuning $\zeta = A - \phi_0 + r\tau'_{\text{eff}}\,\sigma'\mu A^2$ is chosen for optimized power transfer from the pump $F$ to the pumped mode, which makes the expression (*) $A - \zeta + \phi_0 - r\tau'_{\text{eff}}\,\sigma'\mu A^2$ on the right-hand-side of Eq. (10) vanish. With the above-mentioned parameters, Eq. (10) can then be written as by $F = \left(1 + rA + r\tau'_{\text{eff}}\,\sigma' A^2\right)^2 A$, leading to $A \approx 1$ for a large range of technically relevant normalized pump powers $F$ between 1 and 100. Note that the following investigation aims at identifying the dominant terms in Eq. (9) and that the symbol "$\approx$" used is to be understood as an order-of-magnitude quantification rather than as an approximate equality.

To identify the side bands at which MI will occur first, we need to find values of the side-band offset $\Omega$ that maximize the gain rate $\lambda$. To this end, we simplify Eq. (9) by reducing it to its dominant terms, assuming that the offset of the MI-generated side-bands from the pump is of the order of $\Omega \approx 10$. With the above-mentioned parameters, this leads to

$$\lambda(\Omega) = -1 - r\left(2 + \tau'_{\text{eff}}\sigma' A\right)A + \Re\left\{ \sqrt{A^2\left(r^2 + 1\right) - \left(\zeta + d\Omega^2 - 2A + r\tau'_{\text{eff}}\sigma'\mu A^2\right)^2} \right\}, \tag{11}$$

see Appendix B for details. For simplicity, we treat $\Omega$ as a continuous non-negative real-valued variable even though it was originally defined as an integer parameter $\Omega \in \mathbb{N}_+$. The sidebands experiencing the highest gain rate are then obtained from $\zeta + d\Omega^2 - 2A + r\tau'_{\text{eff}}\sigma'\mu A^2 = 0$, leading to

$$\Omega_{\max} = \sqrt{\left(2A - r\tau'_{\text{eff}}\sigma'\mu A^2 - \zeta\right)/d}. \tag{12}$$

Note that Eq. (12) implies an appropriate choice of the detuning $\zeta$ such that $\left(2A - r\tau'_{\text{eff}}\sigma'\mu A^2 - \zeta\right)/d > 0$. Note also that a strict derivation of $\Omega_{\max}$ by computing $\mathrm{d}\lambda/\mathrm{d}\Omega = 0$ will yield $\Omega = 0$ as an additional local extremum, specifically a local maximum for $d\left(\zeta - 2A + r\tau'_{\text{eff}}\sigma'\mu A^2\right) > 0$ and a local minimum for $d\left(\zeta - 2A + r\tau'_{\text{eff}}\sigma'\mu A^2\right) < 0$. This extremum is not considered further, since the associated gain parameter $\lambda(0)$ is always smaller than $\lambda(\Omega_{\max})$, which is given by

$$\lambda(\Omega_{\max}) = -1 - r\left(2 + \tau'_{\text{eff}}\sigma' A\right)A + A\sqrt{r^2+1} = -1 + \left(\sqrt{r^2+1} - 2r\right)A - r\tau'_{\text{eff}}\sigma' A^2. \tag{13}$$

Note that in absence of free carriers, i.e., $\tau'_{\text{eff}}\sigma' = 0$, a relation for $\lambda(\Omega_{\max})$ equivalent to Eq. (13) can be directly obtained from Eq. (9) without any further approximations. Note also that Eq. (13) reproduces the well-known fact that $A > 1$ is a necessary condition for MI to occur in the absence of TPA and FCA. The presence of these effects may either increase the required normalized power $A$ to values larger than 1 or completely inhibit MI. Specifically, $\lambda(\Omega_{\max})$ is negative for any value of $A$ for sufficiently high TPA parameters $r \geq 1/\sqrt{3}$, i.e., MI cannot occur, irrespective of the pump power. On the other hand, positive values of $\lambda(\Omega_{\max})$ may be found for certain normalized powers $A$ if both of the following conditions are satisfied:

$$0 \leq r < 1/\sqrt{3}, \qquad 0 \leq \tau'_{\text{eff}}\,\sigma' < \frac{1}{4r}\left(\sqrt{r^2+1} - 2r\right)^2 . \tag{14}$$

Note that the upper limit for $r$ of $1/\sqrt{3}$ is exact and is in agreement with results also obtained from a bifurcation study of the LLE including TPA [32]. For the values of $r$ and $\tau'_{\text{eff}}\,\sigma'$ specified by Eq. (14), we use Eq. (13) to compute the minimum threshold power $A_{\text{th}}$ of the pumped resonator mode that is required to achieve MI, i.e., $\lambda(\Omega_{\max}) > 0$. The forcing $F_{\text{th}}$ required to achieve $A_{\text{th}}$ is then determined from Eq. (10). For maximizing the power transfer from $F_{\text{th}}$ to $A_{\text{th}}$, the detuning is chosen as $\zeta = A_{\text{th}} - \phi_0 - r\tau'_{\text{eff}}\,\sigma'\mu A_{\text{th}}^2$ by appropriate adjustment of the pump frequency, thus eliminating the expression marked by a star (*) in Eq. (10). For this detuning, maximum gain is found for modes with offset $\Omega_{\max} = \sqrt{\left(A_{\text{th}} - \phi_0\right)/d}$. For anomalous dispersion, $d > 0$ ($\beta_2 < 0$), real-valued $\Omega_{\max}$ can be found as long as $\phi_0 < A_{\text{th}}$, which includes also complete absence of AMC, $\phi_0 = 0$. In contrast to that, normal dispersion, $d < 0$ ($\beta_2 > 0$), requires $\phi_0 > A_{\text{th}}$, i.e., a spectral shift of the resonance caused by sufficiently strong AMC, leading to a real-valued $\Omega_{\max}$. For real $\Omega_{\max}$, Figure 2 displays the color-coded threshold forcing $F_{\text{th}}$ that is required to achieve MI as a function of the normalized TPA coefficient $r$ and the free-carrier influence $\tau'_{\text{eff}}\,\sigma'$. The color-coded map is limited to the ranges within which MI can be achieved, see Eq. (14), while the remainder of the plot is kept in grey. We find that $F_{\text{th}}$ increases continuously with increasing $r$ and $\tau'_{\text{eff}}\,\sigma'$, which is caused by both an increase of $A_{\text{th}}$ needed to achieve positive $\lambda(\Omega_{\max})$ according to Eq. (13), and a reduced power transfer from the pump $F$ to the pumped mode $A$, Eq. (10). In absence of FCA, i.e., $\tau'_{\text{eff}}\,\sigma' = 0$, MI is possible for sufficiently weak TPA $r < 1/\sqrt{3}$, indicated by a vertical dashed line, and the associated threshold forcing $F_{\text{th}}$ diverges for $r \to 1/\sqrt{3}$. This can be seen in Eq. (13), in where the factor $\sqrt{r^2+1} - 2r \to 0$ vanishes for $r \to 1/\sqrt{3}$ and thus $A_{\text{th}} \to \infty$ is needed to achieve positive $\lambda(\Omega_{\max})$ for $\tau'_{\text{eff}}\,\sigma' \to 0$. For $r < 1/\sqrt{3}$ the threshold forcing at the edge of the MI-enabling parameter space remains finite. In Appendix C we check the result of Fig. 2 for a specific set of parameters $\sigma', \mu, d > 0$ by evaluating Eqs. (9) and (10) numerically for varying $r$ and $\tau'_{\text{eff}}$ without the approximations involved in Eq. (13). In this investigation, we again assume that AMC is absent, i.e., $\phi_{\Omega'} = 0\,\forall\,\Omega'$. The relative deviation of the threshold forcing found by the numerical evaluation from its analytically approximated counterpart stays below 1 % as long as the threshold forcing is not significantly larger than 10. We hence conclude that the simplified procedure leading to Fig. 2 can be considered sufficiently accurate.

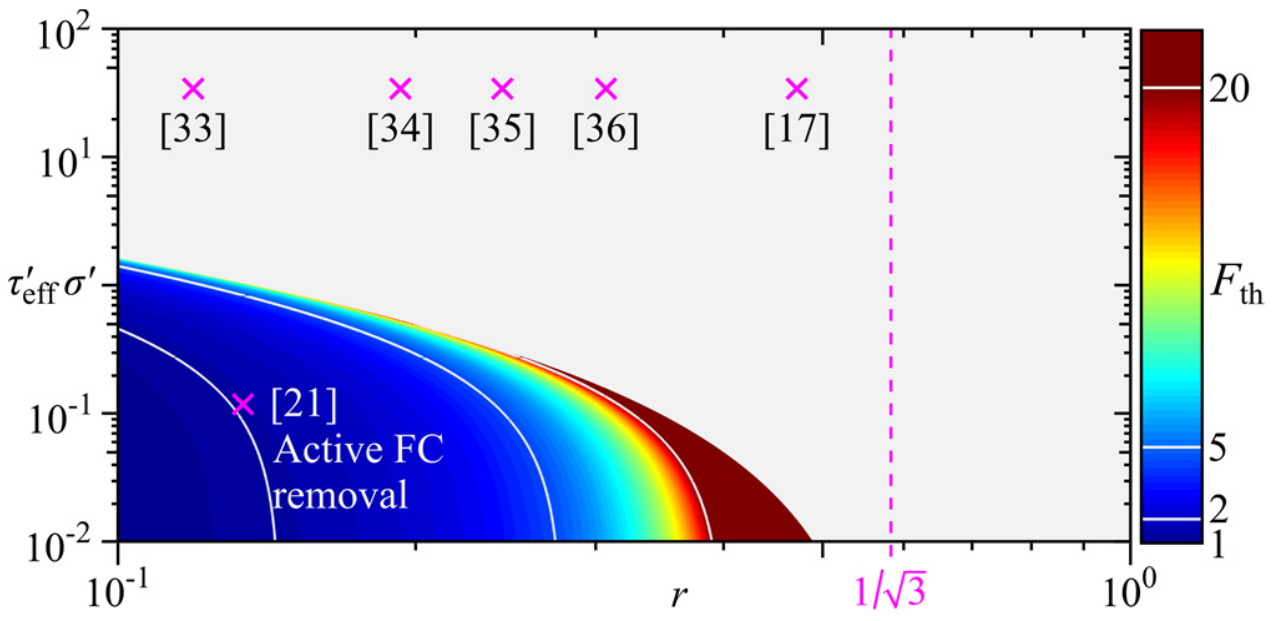

**Figure 2:** Threshold forcing needed to achieve MI as a function of $r$ and $\tau'_{\mathrm{eff}}\sigma'$. Modulation instability and comb formation can be observed only to the left of the vertical dashed line $r = 1/\sqrt{3}$. The grey area indicates the parameter space, for which modulation instability does not occur. Magenta crosses mark real data points reported for silicon-photonic waveguides [17,21,33–36], assuming waveguide losses of $\alpha_i = 46\,\mathrm{m}^{-1}$ ( $2\,\mathrm{dB\,cm}^{-1}$ ) [21]. Reference [21] reports on silicon-photonic waveguides in which free carriers are actively removed by a reverse-biased *p-i-n*-junction, leading to a dwell time of $\tau_{\mathrm{eff}} = 12\,\mathrm{ps}$. In all other cases, no free-carrier removal was used, leading to dwell times of the order of $1 \ldots 3\,\mathrm{ns}$ according to [17,21,37]. Specifically, a value of $3\,\mathrm{ns}$ was used for the data points related to the references [33–36], which do not specify values for $\tau_{\mathrm{eff}}$. The value for the FCA cross-section $\sigma = 1.45\times10^{-21}\,\mathrm{m}^2$ is consistently found in various publications [17,21,34]. The operating wavelength is $1550\,\mathrm{nm}$.

Figure 2 also shows published experimental data (magenta crosses) for the TPA coefficient $r$ and for the product $\tau'_{\mathrm{eff}}\sigma'$ of the FC cross-section and dwell time. The published figures for $r$ are all in the same order of magnitude [17,21,33–36] and stay below the limiting value of $1/\sqrt{3}$ as given in Eq. (14). For simple silicon-photonic waveguides without active free-carrier removal, dwell times $\tau_{\mathrm{eff}}$ are of the order of 1 … 3 ns [17,21,37], thereby clearly inhibiting MI. However, active free-carrier removal by a reverse-biased *p-i-n*-junction can effectively reduce the dwell time to values of, e.g., 12 ps [21], such that modulation instability and frequency comb formation then become possible at telecommunication wavelengths. The lower limit for $\tau'_{\mathrm{eff}}\sigma'$ that is achievable by active free-carrier removal is dictated by the saturation drift velocity $v_{\mathrm{FC}}$ of the free carriers, which is of the order of $10^5\,\mathrm{m\,s}^{-1}$ for electrons in silicon [38]. For a microresonator with a waveguide width of $w = 480\,\mathrm{nm}$ as shown in Fig. 1, the theoretically estimated carrier dwell time can be as small as $\tau_{\mathrm{eff}} = w/v_{\mathrm{FC}} = 4.8\,\mathrm{ps}$ which leads to $\tau'_{\mathrm{eff}}\sigma' = 0.05$ and clearly enables MI.

## IV. SILICON MICRORESONATOR FOR COMB GENERATION AT TELECOM WAVELENGTHS

In Figure 1 we show the structure of a silicon microresonator along with geometrical parameters that lead to anomalous group velocity dispersion around $1550\,\mathrm{nm}$ [24] and to a free spectral range (FSR) of $100\,\mathrm{GHz}$. The geometry of the cross-sectional design for fast carrier removal is similar to the one used in [21]. The waveguide is undoped and is part of the intrinsic zone of a *p-i-n*-junction [23,39]. A reverse voltage applied through vertical interconnect accesses (vias) to the $p^+$-doped (red) and $n^-$-doped (blue) regions of the *p-i-n*-junction leads to efficient removal of free carriers such that dwell times of the order of the round-trip time can be achieved. Finite-element simulations yield a second-order dispersion parameter of $\beta_2 = -587\,\mathrm{ps}^2\,\mathrm{km}^{-1}$, a group refractive index of $n_{\mathrm{g}} = 4.15$, and a nonlinearity parameter of $\gamma = 257\,\mathrm{W}^{-1}\,\mathrm{m}^{-1}$ at $1550\,\mathrm{nm}$, see Appendix D for details. For the TPA and FCA parameters marked by a magenta cross in Fig. 2 [21], the threshold forcing for MI amounts to $F_{\mathrm{th}} = 2.06$. Assuming a power loss coefficient of $\alpha_i = 46\,\mathrm{m}^{-1}$ ( $2\,\mathrm{dB\,cm}^{-1}$ ) and critical coupling, which corresponds to a Q-factor of approximately $2\times10^5$, this threshold forcing translates into a threshold pump power of $P_{\mathrm{in}} \approx 12\,\mathrm{mW}$ measured in the on-chip bus waveguide. Q-factors of the order of $10^5$ have been demonstrated using commercial silicon-photonic foundry processes [40].

## V. SIMULATING THE DYNAMICS OF COMB GENERATION

With a specific resonator design at hand, we next perform a time integration of the LLE to validate our theoretical predictions on comb formation. We start our consideration from the microresonator design described in Section IV, which features a dispersion coefficient $d = 0.0025$, a TPA coefficient $r = 0.133$, an FCA cross-section $\sigma' = 0.0157$, and an FC dispersion coefficient $\mu = 7.5$ [17,41], see Appendix G, Tables G1 and G2 for a list of the underlying physical microresonator parameters along with their connection to the normalized quantities. Forcing $F$, detuning $\zeta$ and FC dwell time $\tau'_{\text{eff}}$ are externally controllable parameters and are varied in our simulations. We use a resolution of $\Delta\tau' = 2\pi/1024$ ($\Delta\tau = 9.7$ fs) for the normalized fast time by dividing the round-trip time into 1024 parts, and we set the slow-time step-size to $\Delta t' = 0.003$ ($\Delta t = 0.9\,\text{ps}$). In each simulation, we perform $100\,000$ time steps for the slow time. The initial field for each simulation is given by 1024 complex numbers with random phases between 0 and $2\pi$ and random amplitudes between 0 and $10^{-14}$. The maximum amplitude of the initial field is chosen such that the power of the initial field is negligible compared to the forcing but can still start the evolution of the differential equation system. More details on the integration of the coupled Eqs. (5) and (6) can be found in Appendix E. We analyze four different cases, see Columns (a)-(d) of Fig. 3 for the results.

In Column (a) (*Inhibited modulation instability*), the parameters are set as follows: Forcing $F = 5$ (pump power $P_{\text{in}} = 30\,\text{mW}$), dwell-time $\tau'_{\text{eff}} = 2\pi \times 5.00$ ($\tau_{\text{eff}} = 50\,\text{ps}$), detuning $\zeta = 2.2$ ($\Delta\omega = 2\pi \times 1.2\,\text{GHz}$), and $d = 0.0025$, corresponding to anomalous dispersion, i.e., $\beta_2 < 0$, see Table 1. The simulation parameters are listed in Row R1 of Fig. 3. The dispersion profile of the 101 central frequency comb modes, represented in normalized terms by $\varphi'(\Omega') = -\Omega'^2 d + \phi_{\Omega'}$, is shown in Row R2 as a function of the mode index $\Omega'$. In physical terms, the dispersion profile $\varphi(\Omega')$ corresponds to the phase deviation accumulated by each comb mode $\Omega'$ over a single round-trip $t_{\text{R}}$ in the resonator due to dispersion and avoided mode crossings. This can be seen by introducing Eq. (1) into a reduced version of Eq. (2), where only the fifth and sixth term $\mathrm{i}\left[-(\beta_2 L/2)\partial^2/\partial\tau^2 + \hat{\Phi}_{\text{AMC}}\right]\underline{E}(t,\tau)$ on the r.h.s. are maintained. Using Eq. (4), and the normalized quantities defined in Table 1 , we obtain

$$\mathrm{i}\left[-\frac{\beta_2 L}{2}\frac{\partial^2}{\partial\tau^2} + \hat{\Phi}_{\text{AMC}}\right]\sum_{\Omega'}\underline{E}_{\Omega'}\mathrm{e}^{\mathrm{i}2\pi\Omega' t/t_{\text{R}}} = \mathrm{i}\sum_{\Omega'}\varphi(\Omega')\underline{E}_{\Omega'}\mathrm{e}^{\mathrm{i}2\pi\Omega' t/t_{\text{R}}},$$

$$\varphi(\Omega') = \frac{\beta_2 L}{2}\left(\Omega' \times \frac{2\pi}{t_{\text{RT}}}\right)^2 + \delta\omega_{\Omega'} t_{\text{RT}}, \tag{15}$$

$$\varphi'(\Omega') = \varphi(\Omega')\frac{n_{\text{g}}}{\alpha_i t_{\text{RT}} c} = -\Omega'^2 d + \phi_{\Omega'}$$

The above-mentioned choice of forcing and dwell time ensures that the normalized gain rate $\lambda(\Omega)$ (with $\Omega = |\Omega'|$), Eqs. (9) and (10), is always negative, see Row R3. As a consequence, modulation instability cannot occur and the only mode with non-zero power is the pumped mode at modal index $\Omega' = 0$. In Row R4, the color-coded power spectrum $\left|\underline{\hat{a}}_{\Omega'}(t')\right|^2$ is shown as a function of the normalized (slow) time $t'$ in the range $0 \le t' \le 300$ and of the modal index $\Omega'$ in the range $-50 \le \Omega' \le +50$.

Row R5 displays the final power spectrum at $t' = 300$ as a function of the modal index $\Omega'$. In Row R6, the color-coded intra-cavity (IC) power $\left|\underline{a}(t',\tau')\right|^2$ is depicted as a function of normalized slow time $t'$ and normalized fast (retarded) time $\tau'$. In absence of modulation instability, the IC power remains constant along the circumference of the resonator, $\left|\underline{a}(t',\tau')\right|^2 = 1.13$, see Rows R6 and R7.

In Column (b) (*Modulation instability*), we keep all parameters of Column (a) except for $\tau'_{\mathrm{eff}}$, which is reduced to $2\pi \times 1.22$. This corresponds to a physical dwell time $\tau_{\mathrm{eff}} = 12.2\,\mathrm{ps}$, which has previously been demonstrated in a comparable silicon-photonic waveguide [21]. In this case, the computed normalized gain rate is positive within a certain range of modal indices $\Omega'$, see Fig. 3(b) Row R3 (red dots). During the evolution of the power spectrum with slow time $t'$, first sidebands emerge near $t' = 50$, Row R4. A comparison with the gain rate plot in Row R3 shows that the positions $\Omega' \approx 26$ of the initial sidebands coincide with the maxima of the gain parameter $\lambda(\Omega')$. A power spectrum at $t' = 300$ is to be seen in Row R5. The phase-locked modes lead to a regular temporal pulse pattern which is visible in Rows R6 and R7. Because the spectral distance of the dominating modes in Row R5 is much larger than the resonator's FSR, the resulting pulse period is much smaller than $f_{\mathrm{FSR}}^{-1}$. Note that there is a slow drift of this pulse train within the retarded time frame $\tau'$, Row R6. This drift is attributed to the fact that the retarded time frame is defined via the group velocity $v_{\mathrm{g}}$ of the "cold" resonator. In the presence of free carriers in the "hot" resonator, FC dispersion increases the actual group velocity and leads to a residual time shift, which accumulates over subsequent round-trips and therefore grows continuously with slow time $t'$. Mathematically, this behavior can be also seen from the terms $\exp(\pm \mathrm{i}\lambda_i t')\exp(\pm \mathrm{i}\Omega\tau') = \exp\left(\pm \mathrm{i}\Omega\left(\lambda_i t'/\Omega + \tau'\right)\right)$ in Eq. (8), which describe a time shift of the optical field $\underline{a}(t',\tau')$ that continuously increases with slow time $t'$ within the fast time scale.

In Column (c) (*Dissipative Kerr soliton generation*), we increase the forcing to $F = 8$ ($50\,\mathrm{mW}$). The detuning is kept constant at $\zeta = 3$ until $t' = 90$, then increased linearly to $\zeta = 3.8$ until $t' = 150$, and then kept constant again until the end of the simulation, see Row R1 of Fig. 3(c). Such a procedure allows generating single-soliton states [42]. In Row R3, the gain rate is depicted for $t' = 0$, i.e., before the detuning sweep, showing a broad range of modes that experience parametric gain. In the simulated slow-time evolution of the spectrum, Row R4, the first sidebands become visible around $t' \approx 70$, and the spectral position of these sidebands coincides with the maxima of computed gain rate in Row 3. The maximum of the gain parameter in Column (c) is slightly smaller than the one obtained for the scenario described in Column (b), and thus the sidebands become only visible at a later normalized time *t*'. The final power spectrum obtained at the end of the simulation, Row R5, is a very regular frequency comb with a smooth envelope, which is typical for a single dissipative Kerr soliton [9] circulating in the ring. The evolution of the color-coded intracavity power, shown in Row R6, reveals the emergence of multiple pulses at modulation instability onset around $t' \approx 70$. Due to the swept detuning, most of the pulses diminish over time, which is an experimentally well observed behavior [42]. Note that, while increasing the detuning, the pulses disappear consecutively in direction of decreasing $\tau'$, starting to the left of the finally remaining pulse, Row R6. This is caused by the fact that, within a sequence of pulses, the last pulse is always subject to the highest accumulated FC concentration and thus experiences that highest FCA loss. The final value of $\zeta$ was chosen such that a single pulse remains in the cavity, which can be seen from the plot of the IC power $\left|a(t' = 300, \tau')\right|^2$ in Row R7. In this simulation, FCA causes an even more pronounced change of the group velocity compared to Fig. 3(b) R6, leading to a stronger temporal shift of the soliton pulse within the retarded time frame $\tau'$, while the spectrum remains constant.

In Column (d) (*Modulation instability in a microresonator with normal dispersion*), the forcing is set back to $F = 5$, and the anomalous dispersion profile used in Fig 3(a)-(c) is inverted to obtain normal group-velocity dispersion, $d = -0.0025$, see Row R1 for all parameters. Additionally, we introduce an avoided mode crossing (AMC) which causes phase shifts $\phi_{\Omega'}$ that disturb the dispersion profile. The AMC arises due to a coupling of resonator modes with similar resonance frequency, but different transverse field distributions. The two coupled modes belong to different mode families, characterized by their respective free spectral range. Representing the dispersion profile of the second waveguide mode family in the dispersion diagram of the first mode family leads to equidistant points on an approximately straight line, which is indicated in red in Row R2 of Column (d). The resonance frequencies of the coupled modes are indicated by blue filled circles connected by grey lines, which deviate from the resonance frequencies of the unperturbed modes $\omega_{\Omega'}$ by $\delta\omega_{\Omega'}$, see Appendix F and Ref. [43] for details. The FSR of the second mode family and the coupling strength of the two transverse modes are chosen such that the strongest resonance shift amounts to $\delta\omega_{-10} = -2\pi \times 630\,\text{MHz}$ ($\phi_{-10} = -1.19$, $\phi_0 \approx 0.06$, $\phi_{+10} \approx 0$). This corresponds to a local 1% change of the FSR, which is the same order of magnitude as reported for experimentally investigated microresonators [28,43]. The phase shifts $\phi_{\Omega'}$ induced by the avoided mode crossing alter the gain parameter such that it becomes positive for certain sidebands, with a gain maximum at sidebands $\Omega' = \pm 10$, see Row R3. In the evolution of the power spectrum, Row R4, comb lines emerge at these positions. For better visibility, the comb line at $\Omega' = 10$ is framed by a broken line and horizontally enlarged. The stationary power spectrum in Row R5 shows an asymmetry. Compared to the previously considered scenarios, the IC power distribution exhibits a stronger temporal shift of the soliton pulse within the retarded within the retarded time frame $\tau'$, see the zoom-in in Row R6. This is caused by a contribution of the AMC-induced phase shifts to the imaginary part of the gain rate, see Eq. (9). Apart from this time shift, the final power distribution inside the resonator exhibits a stationary regular pattern, Row R7.

Note that silicon-photonic waveguides with anomalous group-velocity dispersion around $1550\,\text{nm}$ need a careful design and can only be achieved in a limited parameter space of widths $w$ and heights $h$ [24]. To ensure the technical relevance of the scenarios investigated in Columns (a), (b), and (c) of Fig. 3, we derived the normalized dispersion parameter along with the corresponding nonlinearity parameter from a specific waveguide design ($w = 480\,\text{nm}$, $h = 300\,\text{nm}$), see Table 1 and Appendix D for details. In contrast to that, normal dispersion can be achieved for a rather large parameter range of waveguide widths and heights, including rather large cross-sections for which multi-mode propagation and avoided mode-crossings may occur [24]. Silicon-photonic microresonators corresponding to the scenario considered in Column (d) may therefore by realized for multiple different waveguide geometries.

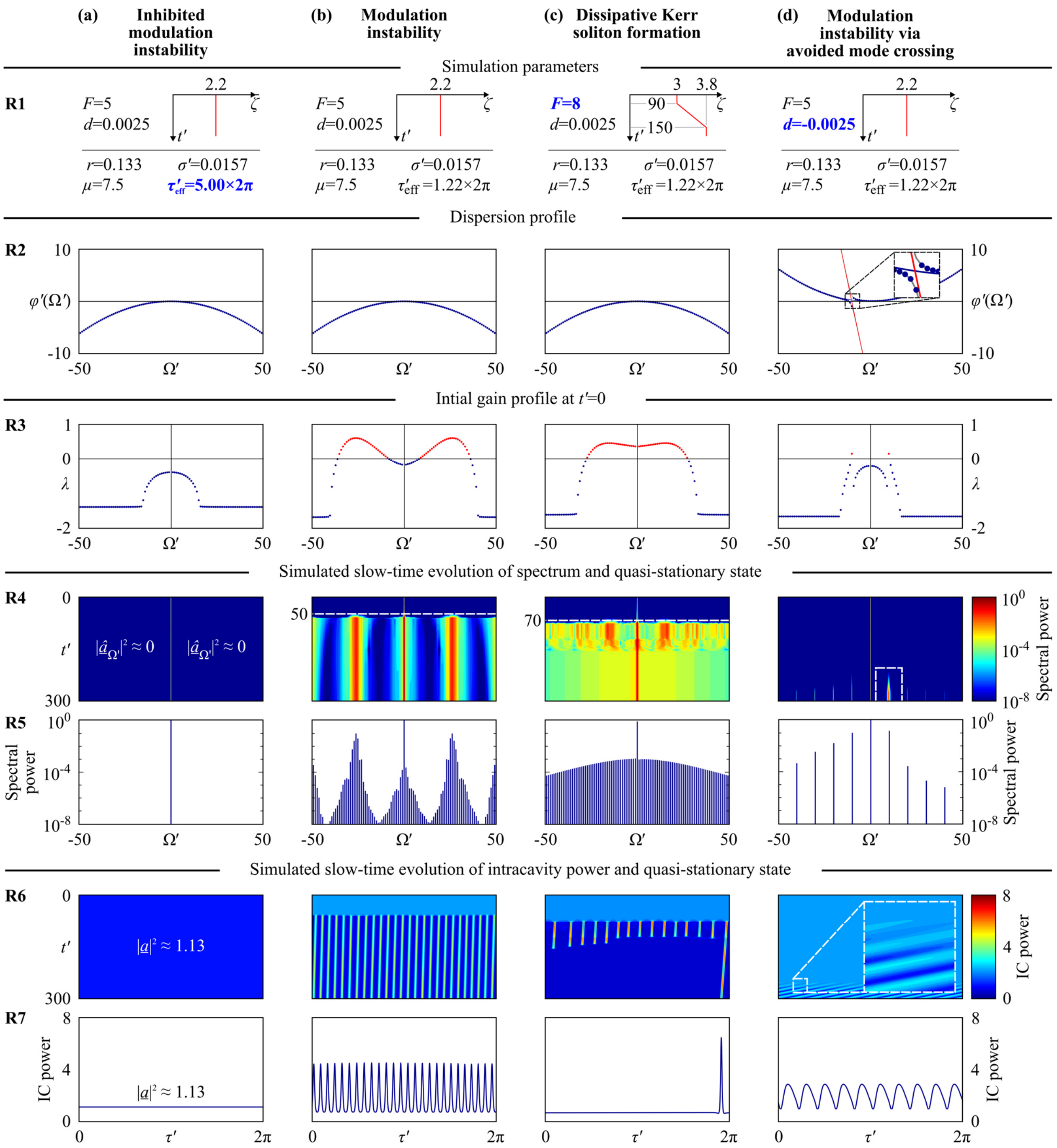

**Figure 3:** Results of time integration of the LLE for different resonator designs and pump parameters. Row R1: Simulation parameters. Row R2: Dispersion profile for the central 101 modes. Row R3: Computed normalized gain rates at the beginning of the simulation (slow time $t' = 0$). Row R4: Evolution of power spectra with normalized slow time $t'$. Row R5: Spectra at stationary state for $t' = 300$. Row R6: Color-coded intracavity (IC) power $\left|\underline{a}\left(t',\tau'\right)\right|^2$ evolving during comb formation over $t'$ as a function of the fast time $\tau'$. We observe a temporal shift of the IC pulse pattern within the retarded time frame of the fast time axis $\tau'$, which can, e.g., be caused by FC dispersion or avoided mode crossings. Besides this shift along the $\tau'$-axis, the IC power distribution evolves into a "quasi-stationary" final state in all cases. Row R7: Quasi-stationary IC power $\left|\underline{a}\left(t' = 300, \tau'\right)\right|^2$. Column (a): Modulation instability inhibited by long free carrier dwell time $\tau_{\mathrm{eff}}$, which corresponds to five times the cavity round-trip time $t_{\mathrm{RT}}$. Column (b): Modulation instability with sufficiently reduced carrier dwell time $\tau_{\mathrm{eff}} = 1.22 \times t_{\mathrm{RT}}$. Column (c): Single soliton formation for dynamically increasing detuning. Column (d): Comb formation in a normal-dispersion resonator with avoided mode crossing.

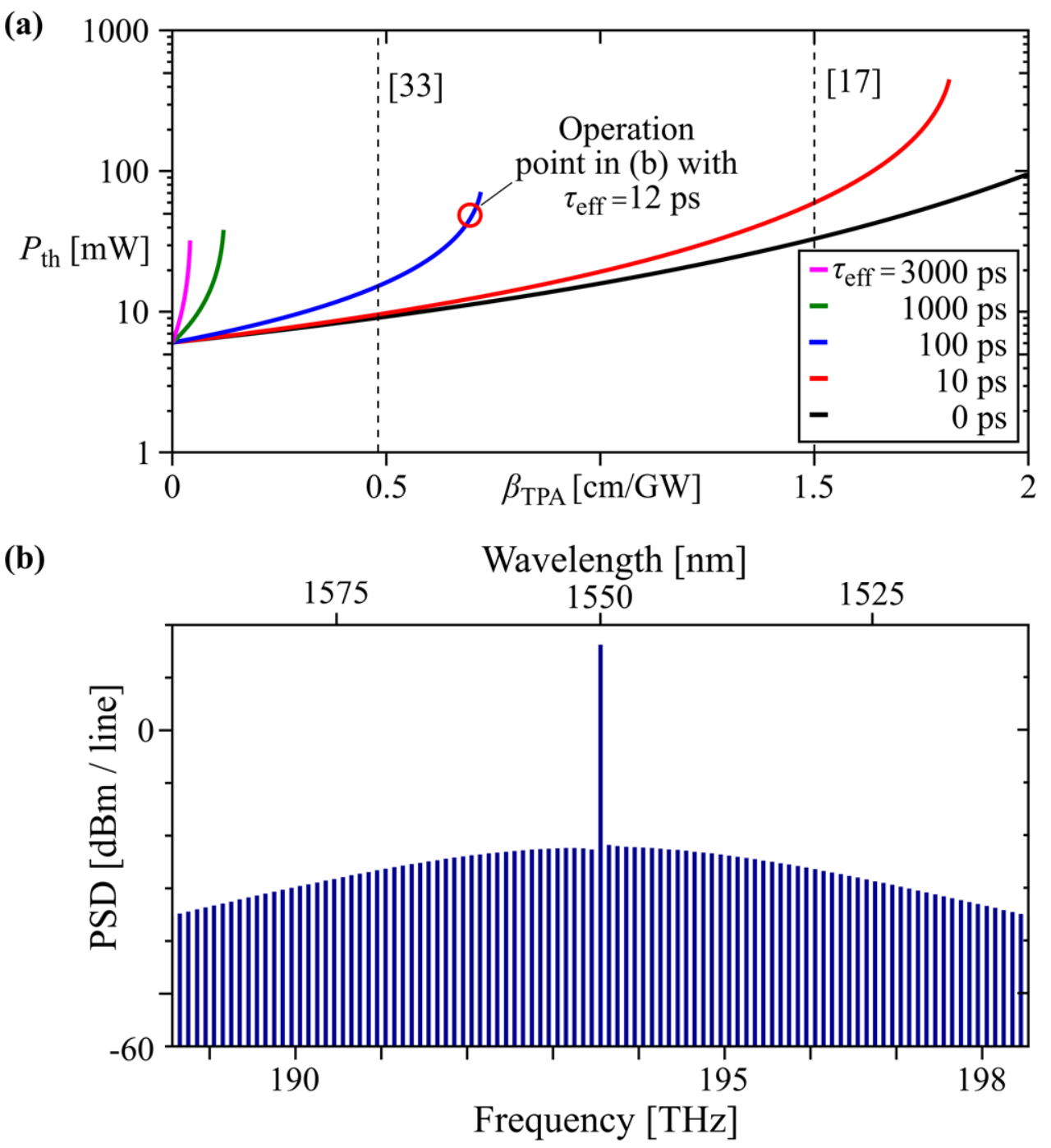

**Figure 4:** (a) On-chip threshold pump power for the onset of modulation instability as a function of TPA coefficient $\beta_{TPA}$ for different FC dwell times $\tau_{eff}$. We assume a silicon microresonator with an $f_{FSR} = 100\,GHz$ at $1550\,nm$, see Section IV for details. The microresonator is assumed to have $2\,dB\,cm^{-1}$ waveguide loss, anomalous dispersion and critical coupling to the bus waveguide. Reported values of $\beta_{TPA}$ for silicon range from $0.45\,cm\,GW^{-1}$ to $1.5\,cm\,GW^{-1}$ as indicated by the vertical dashed lines. (b) Soliton comb spectrum obtained in the bus waveguide after the microresonator. The resonator is identical to the one considered in Subfigure (a). We assume $\beta_{TPA} = 0.7\,cm\,GW^{-1}$, $\tau_{eff} = 12ps$, and $P = 50\,mW$ as indicated by a red circle in Subfigure (a). The spectrum is derived from the normalized intra-cavity comb spectrum indicated in Fig. 3(c), Row R5.

## VI. DISCUSSION AND CONCLUSIONS

The simulation results indicate that modulation instability in silicon-photonic microresonators at telecommunication wavelengths is most likely to be observed if the waveguide is designed for anomalous group-velocity dispersion, if FCA is mitigated by a reverse-biased *p-i-n*-junction which leads to a sufficiently small carrier dwell time, and if the pump power and detuning are chosen properly. If an avoided mode crossing induces local dispersion shifts, microresonators with otherwise normal dispersion can also exhibit modulation instability. In both cases, moderate on-chip pump powers in the range of $10\,mW$ to $50\,mW$ are sufficient to initiate modulation instability and to generate frequency combs. The required pump powers depend strongly on the actual values of the TPA coefficient $\beta_{TPA}$, which is expressed by the normalized TPA parameter $r$, and on the actual FC dwell time $\tau_{eff}$ and its normalized counterpart $\tau'_{eff}$. We illustrate this dependence in Fig. 4 (a), where the on-chip threshold pump power $P_{th}$ for modulation instability is displayed as a function of the TPA coefficient with the FC dwell time as a parameter. We assume critical coupling, a resonator design and waveguide properties as described in Section IV, a Kerr coefficient of $n_2 = 6.5\times10^{-18}\,m^2\,W^{-1}$, and a pump wavelength of $1550\,nm$. We find that $\tau_{eff}$ has to be of the order of $100\,ps$ or less to enable comb formation across the range of reported values for $\beta_{TPA}$ in silicon, which reach from $0.45\,cm\,GW^{-1}$ [33] to $1.5\,cm\,GW^{-1}$ [17] as indicated by vertical dashed lines in Fig. 4(a). For larger dwell times, FCA prevents comb formation, which is indicated by the fact that the green and magenta lines in Fig. 4(a) do not enter the range of reported values for $\beta_{TPA}$.

For $\tau_{\text{eff}} = 10\,\text{ps}$ (red), the threshold pump power varies between $10\,\text{mW}$ and $60\,\text{mW}$, close to the threshold powers in absence of FCA ( $\tau_{\text{eff}} = 0\,\text{ps}$ , black curve) within the range of reported values of $\beta_{\text{TPA}}$ .

For illustration, we calculated the physical frequency comb spectrum that is obtained in the bus waveguide after the microresonator, see Fig. 4(b). We again assume a resonator design and waveguide properties as specified in Section IV, along with a Kerr coefficient of $n_2 = 6.5\times10^{-18}\,\text{m}^2\,\text{W}^{-1}$ and a pump wavelength of $1550\,\text{nm}$ . The TPA coefficient, the effective carrier dwell time, and the pump power are chosen as $0.7\,\text{cm}\,\text{GW}^{-1}$ [21], $12\,\text{ps}$ [21] and $50\,\text{mW}$ , respectively, indicated by a red circle in Fig. 4(a). The physical spectrum shown in Fig. 4(b) is derived from the normalized intra-cavity comb spectrum indicated in Fig. 3(c), Row R5.

Our findings suggest that silicon microresonators may indeed be a viable option for comb generation also at telecommunication wavelengths. We expect that additional effects such as higher-order dispersion, Raman shift [44] and self-steepening may lead to minor corrections of the quantitative predictions, but should not change the qualitative behaviour. More accurate data for $\beta_{\text{TPA}}$ and achievable ranges of $\tau_{\text{eff}}$ will also help to obtain a more precise estimate of threshold powers. Moreover, several approaches may allow to reduce the predicted pump powers. Further improvements in waveguide fabrication may allow to reduce the linear propagation losses to values of, e.g., $0.4\,\text{dB}\,\text{cm}^{-1}$ [45], which is well below the $2\,\text{dB}\,\text{cm}^{-1}$ assumed in this work, but still far above the intrinsic absorption of silicon of less than $0.01\,\text{dB}\,\text{cm}^{-1}$ at telecommunication wavelengths [46]. As an alternative or an addition to reverse-biased *p-i-n* junctions, silicon self-ion implantation may be used to reduce of the FC dwell time at the expense of slightly increased waveguide losses [47]. Alternative waveguide concepts, e. g., silicon-organic-hybrid (SOH) waveguides [26,27] with a high Kerr nonlinearity of order of $\gamma = 100\,\text{W}^{-1}\,\text{m}^{-1}$ and low normalized TPA absorption coefficients of $r \approx 0.036$ , corresponding to $\beta_{\text{TPA}} = 0.5\,\text{cm}\,\text{GW}^{-1}$ for $n_2 = 1.7\times10^{-17}\,\text{m}^2\,\text{W}^{-1}$ [26] may also be used in a microresonator.

In summary, we have presented a theoretical analysis of the impact of nonlinear loss mechanisms such as TPA and FCA on Kerr comb formation. We derive the maximum two-photon absorption and free-carrier lifetime that still permits to achieve modulation instability at sufficiently low pump powers and that can thus lead to frequency comb formation. We show that silicon microresonators are not necessarily unsuited for Kerr comb generation at NIR telecommunication wavelengths, provided that the dwell time of the free-carriers in the waveguide core is reduced by a reverse-biased *p-i-n*-junction and that the pump parameters are chosen appropriately, and we present a specific design of a silicon microresonator with anomalous group-velocity dispersion that may even support formation of dissipative Kerr solitons. A numerical solution of the Lugiato-Lefever equation shows the onset of comb formation in agreement with our theoretical small-signal analysis. Both the numerical and the theoretical small-signal analysis demonstrate that modulation instability can also occur for normal-dispersion resonators if an avoided mode crossing comes into play. This would permit Kerr comb generation in silicon ring resonators with a standard waveguide height of $220\,\text{nm}$ . Our analysis is based on normalized quantities and can thus be applied to wide range of resonator design, materials, and operation conditions. For a specific resonator example, the threshold powers for the onset of modulation instability are extracted, Fig. 4. These data help in designing microresonators for Kerr comb generation on different material platforms with a variety of geometrical parameters.

**ACKNOWLEDGEMENTS**

Funded by the Deutsche Forschungsgemeinschaft (DFG, German Research Foundation) – Project-ID 258734477 – SFB 1173; European Union's Horizon 2020 research and innovation programme grant agreement No. 863322 (TeraSlice), European Research Council (ERC Consolidator Grant 'TeraSHAPE', # 773248); Erasmus Mundus doctorate program Europhotonics (grant number 159224-1-2009-1-FR-ERA MUNDUS-EMJD). Datasets are available from the corresponding authors upon reasonable request.

## APPENDIX

### A. LUGIATO-LEFEVER EQUATION FOR MODELING MODULATION INSTABILITY

When pumping a nonlinear system with a sinusoidal waveform of constant amplitude, spectral sidebands can develop, and the pump amplitude appears modulated. This so-called modulation instability (MI) is the starting point of frequency comb formation in a pumped Kerr-nonlinear optical microresonator. We determine the onset of MI in the presence of two-photon absorption (TPA) and free carrier absorption (FCA) analytically by a small-signal approximation of the modified Lugiato-Lefever equation (LLE). In the following analysis, we use normalized quantities, see Table 1 of the main manuscript. The complex optical field amplitude $\underline{E}(t,\tau)$, Eq. (1) in the main manuscript, is defined with respect to the pump frequency $\omega_\mathrm{P}$. It depends on a "slow" time variable $t = m\,t_\mathrm{RT}$ expressed in multiples $m$ of the fixed round-trip time $t_\mathrm{RT} = L / v_\mathrm{g}$ for a given resonator perimeter $L$ and a group velocity $v_\mathrm{g}$ at the pump frequency $\omega_\mathrm{P}$, as well as on a "fast" time $\tau = t - z / v_\mathrm{g}$, defined by a retarded time frame that depends on the local position $z$ inside the resonator. The corresponding normalized optical field according to Table 1 is $\underline{a}(t',\tau')$, where the slow normalized time variable $t'$ is assumed to be continuous and where the short normalized time variable $\tau'$ defines the position within the cavity, $\tau' \in [0, 2\pi)$. The resonator is driven by a normalized pump field amplitude $\sqrt{F}$. Due to TPA, the optical field generates free carriers (FC) with a normalized density $N'_\mathrm{c}$. Each rapidly varying time-harmonic component $\mathrm{e}^{\mathrm{i}\omega_{\Omega'}\tau}$, $\omega_{\Omega'} = \Omega' \times (2\pi/t_R)$, $\Omega' = 0, \pm1, \pm2, \pm3, \ldots$ in Eq. (1) of the main manuscript is periodic with the round-trip time, leading to periodic boundary conditions $\underline{a}(t',\tau') = \underline{a}(t',\tau' + 2\pi)$ and $N'_\mathrm{c}(t',\tau') = N'_\mathrm{c}(t',\tau' + 2\pi)$ for the complex field amplitude and the carrier density. The normalized optical field is represented as a series $\underline{a}(t',\tau') = \sum_{\Omega'} \underline{a}_{\Omega'}(t')\,\mathrm{e}^{\mathrm{i}\Omega'\tau'}$. The normalized LLE and the differential equation describing the evolution of the free-carrier density ("free-carrier equation") are given in Eqs. (5), (6) in the main manuscript,

$$\frac{\partial \underline{a}(t',\tau')}{\partial t'} = \sqrt{F} + \left[ -1 - \mathrm{i}\zeta + \mathrm{i}\,d\frac{\partial^2}{\partial \tau'^2} + \mathrm{i}\hat{\Phi}'_\mathrm{AMC} + (\mathrm{i} - r)\left|\underline{a}(t',\tau')\right|^2 - \sigma'(1 + \mathrm{i}\mu)N'_\mathrm{c}(t',\tau') \right] \underline{a}(t',\tau'), \tag{A1}$$

$$\frac{\partial N'_\mathrm{c}(t',\tau')}{\partial \tau'} = r\left|\underline{a}(t',\tau')\right|^4 - \frac{N'_\mathrm{c}(t',\tau')}{\tau'_\mathrm{eff}}. \tag{A2}$$

The normalized difference between the pump frequency $\omega_\mathrm{P}$ and the closest resonance frequency $\omega_\mathrm{R}$ of un-pumped resonator is $\zeta$. The second-order dispersion of the microresonator is considered through the normalized dispersion parameter $d$. The quantity $r$ denotes the normalized TPA-coefficient. For the TPA-generated FC, the quantities $\sigma', \mu, \tau'_\mathrm{eff}$ define the normalized absorption cross-section, the normalized contribution to the refractive index, and the normalized dwell-time inside the microresonator waveguide, respectively. The relationship between normalized parameters and physical quantities are summarized in Table 1 of the main manuscript. We further include an operator $\hat{\Phi}'_\mathrm{AMC}$ describing phase shifts $\phi_{\Omega'}$ experienced by individual components $\underline{a}_{\Omega'}(t')$ of the optical field for frequencies $\Omega'$. These phase shifts take into account local resonance frequency shifts caused by mode coupling of different transverse modes, an effect known as avoided mode crossing [43]. The operator can be written as

$$\hat{\Phi}'_{\mathrm{AMC}}\underline{a}(t',\tau') = \sum_{\tilde{\Omega}} \frac{\phi_{\tilde{\Omega}}}{2\pi}\int_0^{2\pi} \underline{a}(t',\tau_1)\exp\left[-\mathrm{i}\tilde{\Omega}\tau_1\right]\mathrm{d}\tau_1 \exp\left[\mathrm{i}\tilde{\Omega}\tau'\right]. \tag{A3}$$

In our simulations shown in the main manuscript, the strength of the mode coupling, which defines the exact values $\phi_{\Omega'}$, was chosen to reflect typical experimental results.

To investigate the condition for the onset of MI in Eq. (A1) the periodicity of $\underline{a}$ and $N'_{\mathrm{c}}$ is exploited to determine the solution for Eq. (A2),

$$N'_{\mathrm{c}}(t',\tau') = \frac{r}{\exp\left[\dfrac{\tau'}{\tau'_{\mathrm{eff}}}\right] - \exp\left[\dfrac{\tau'-2\pi}{\tau'_{\mathrm{eff}}}\right]}\int_{\tau'-2\pi}^{\tau'} \left|\underline{a}(t',\tau_1)\right|^4 \exp\left[\frac{\tau_1}{\tau'_{\mathrm{eff}}}\right]\mathrm{d}\tau_1. \tag{A4}$$

Next, we a assume that the optical field consists of three waves [18]. Specifically, we assume a constant pumped mode $\underline{a}_0$ and two small sidebands $\underline{a}_{\pm\Omega}$ ($\Omega = 1,2,3,\ldots$), the temporal evolution of which is described by complex gain parameter $\underline{\lambda} = \lambda + \mathrm{i}\lambda_i$,

$$\underline{a}(t',\tau') = \underline{a}_0 + \underline{a}_{+\Omega} + \underline{a}_{-\Omega} = \underline{a}_0 + \hat{\underline{a}}_{+\Omega}\mathrm{e}^{\lambda t'}\mathrm{e}^{\mathrm{i}\lambda_i t'}\mathrm{e}^{\mathrm{i}\Omega\tau'} + \hat{\underline{a}}_{-\Omega}\mathrm{e}^{\lambda t'}\mathrm{e}^{-\mathrm{i}\lambda_i t'}\mathrm{e}^{-\mathrm{i}\Omega\tau'}, \qquad \left|\hat{\underline{a}}_{\pm\Omega}\right| << \left|\underline{a}_0\right|. \tag{A5}$$

First, we determine an explicit expression for the free-carrier density $N'_{\mathrm{c}}(t',\tau')$ introduced in Eq. (A4). We apply a small-signal approximation $\left|\underline{a}(t',\tau')\right|^4 \approx \left|\underline{a}_0\right|^4 + 2\left|\underline{a}_0\right|^2\left(\underline{a}_0^*\underline{a}_{+\Omega} + \underline{a}_0\underline{a}_{-\Omega}^*\right) + 2\left|\underline{a}_0\right|^2\left(\underline{a}_0^*\underline{a}_{-\Omega} + \underline{a}_0\underline{a}_{+\Omega}^*\right)$, where the star (*) denotes the complex conjugate. Keeping only terms up to linear order in $\hat{\underline{a}}_{\pm\Omega}$ or $\hat{\underline{a}}_{\pm\Omega}^*$ and solving the integral in Eq. (A4) leads to

$$N'_{\mathrm{c}}(t',\tau') = r\,\tau'_{\mathrm{eff}}\left|\underline{a}_0\right|^2\left[\left|\underline{a}_0\right|^2 + 2\frac{-\mathrm{i}\Omega\tau'_{\mathrm{eff}}+1}{(\Omega\tau'_{\mathrm{eff}})^2+1}\left(\underline{a}_0^*\underline{a}_{+\Omega} + \underline{a}_0\underline{a}_{-\Omega}^*\right) + 2\frac{\mathrm{i}\Omega\tau'_{\mathrm{eff}}+1}{(\Omega\tau'_{\mathrm{eff}})^2+1}\left(\underline{a}_0^*\underline{a}_{-\Omega} + \underline{a}_0\underline{a}_{+\Omega}^*\right)\right]. \tag{A6}$$

This result is now substituted in Eq. (A1) along with the ansatz for $\underline{a}(t',\tau')$, Eq. (A5). Again, we keep only terms up to linear order in $\hat{\underline{a}}_{\pm\Omega}$ or $\hat{\underline{a}}_{\pm\Omega}^*$. We separate non-oscillating terms from terms oscillating with $\mathrm{e}^{\mathrm{i}\Omega\tau'}$ and $\mathrm{e}^{-\mathrm{i}\Omega\tau'}$, and obtain

$$0 = \sqrt{F} + \left[-1 - \mathrm{i}\zeta + \mathrm{i}\phi_0 + (\mathrm{i}-r)\left|\underline{a}_0\right|^2 - \sigma'(1+\mathrm{i}\mu)r\tau'_{\mathrm{eff}}\left|\underline{a}_0\right|^4\right]\underline{a}_0, \tag{A7}$$

$$\begin{aligned}\underline{\lambda}\,\underline{a}_{+\Omega} &= \left[-1-\mathrm{i}\zeta-\mathrm{i}d\Omega^2+\mathrm{i}\phi_{+\Omega}\right]\underline{a}_{+\Omega} + (\mathrm{i}-r)\left[2\left|\underline{a}_0\right|^2\underline{a}_{+\Omega} + (\underline{a}_0)^2\underline{a}_{-\Omega}^*\right] \\ &\quad -(1+\mathrm{i}\mu)\frac{\sigma' r\tau'_{\mathrm{eff}}\left|\underline{a}_0\right|^2}{(\Omega\tau'_{\mathrm{eff}})^2+1}\left[\left((\Omega\tau'_{\mathrm{eff}})^2+3\right)\left|\underline{a}_0\right|^2\underline{a}_{+\Omega} + 2(\underline{a}_0)^2\underline{a}_{-\Omega}^* - \mathrm{i}2\Omega\tau'_{\mathrm{eff}}\left(\left|\underline{a}_0\right|^2\underline{a}_{+\Omega} + (\underline{a}_0)^2\underline{a}_{-\Omega}^*\right)\right],\end{aligned} \tag{A8}$$

$$\begin{aligned}\underline{\lambda}^*\underline{a}_{-\Omega} &= \left[-1-\mathrm{i}\zeta-\mathrm{i}d\Omega^2+\mathrm{i}\phi_{-\Omega}\right]\underline{a}_{-\Omega} + (\mathrm{i}-r)\left[2\left|\underline{a}_0\right|^2\underline{a}_{-\Omega} + (\underline{a}_0)^2\underline{a}_{+\Omega}^*\right] \\ &\quad -(1+\mathrm{i}\mu)\frac{\sigma' r\tau'_{\mathrm{eff}}\left|\underline{a}_0\right|^2}{(\Omega\tau'_{\mathrm{eff}})^2+1}\left[\left((\Omega\tau'_{\mathrm{eff}})^2+3\right)\left|\underline{a}_0\right|^2\underline{a}_{-\Omega} + 2(\underline{a}_0)^2\underline{a}_{+\Omega}^* + \mathrm{i}2\Omega\tau'_{\mathrm{eff}}\left(\left|\underline{a}_0\right|^2\underline{a}_{-\Omega} + (\underline{a}_0)^2\underline{a}_{+\Omega}^*\right)\right].\end{aligned} \tag{A9}$$

We divide Eq. (A8) by $\mathrm{e}^{\lambda t'}\mathrm{e}^{\mathrm{i}\lambda_i t'}\mathrm{e}^{\mathrm{i}\Omega\tau'}$ and Eq. (A9) by $\mathrm{e}^{\lambda t'}\mathrm{e}^{-\mathrm{i}\lambda_i t'}\mathrm{e}^{-\mathrm{i}\Omega\tau'}$, and we introduce $\bar{\phi} = (\phi_{+\Omega}+\phi_{-\Omega})/2$ and $\Delta\phi = (\phi_{+\Omega}-\phi_{-\Omega})/2$. Both equations are then expressed in terms of $\hat{\underline{a}}_{\pm\Omega}$ instead in terms of $\underline{a}_{\pm\Omega}$,

$$\underline{\lambda}\underline{\hat{a}}_{+\Omega} = \left[-1-\mathrm{i}\zeta-\mathrm{i}d\,\Omega^2+\mathrm{i}\overline{\phi}+\mathrm{i}\Delta\phi\right]\underline{\hat{a}}_{+\Omega}+\left(\mathrm{i}-r\right)\left[2\left|\underline{a}_0\right|^2\underline{\hat{a}}_{+\Omega}+\left(\underline{a}_0\right)^2\underline{\hat{a}}^*_{-\Omega}\right]$$
$$-\left(1+\mathrm{i}\mu\right)\frac{\sigma' r\,\tau'_{\mathrm{eff}}\left|\underline{a}_0\right|^2}{\left(\Omega\tau'_{\mathrm{eff}}\right)^2+1}\left[\left(\left(\Omega\tau'_{\mathrm{eff}}\right)^2+3\right)\left|\underline{a}_0\right|^2\underline{\hat{a}}_{+\Omega}+2\left(\underline{a}_0\right)^2\underline{\hat{a}}^*_{-\Omega}-\mathrm{i}2\Omega\tau'_{\mathrm{eff}}\left(\left|\underline{a}_0\right|^2\underline{\hat{a}}_{+\Omega}+\left(\underline{a}_0\right)^2\underline{\hat{a}}^*_{-\Omega}\right)\right], \quad \text{(A10)}$$

$$\underline{\lambda}^*\underline{\hat{a}}_{-\Omega} = \left[-1-\mathrm{i}\zeta-\mathrm{i}d\,\Omega^2+\mathrm{i}\overline{\phi}-\mathrm{i}\Delta\phi\right]\underline{\hat{a}}_{-\Omega}+\left(\mathrm{i}-r\right)\left[2\left|\underline{a}_0\right|^2\underline{\hat{a}}_{-\Omega}+\left(\underline{a}_0\right)^2\underline{\hat{a}}^*_{+\Omega}\right]$$
$$-\left(1+\mathrm{i}\mu\right)\frac{\sigma' r\,\tau'_{\mathrm{eff}}\left|\underline{a}_0\right|^2}{\left(\Omega\tau'_{\mathrm{eff}}\right)^2+1}\left[\left(\left(\Omega\tau'_{\mathrm{eff}}\right)^2+3\right)\left|\underline{a}_0\right|^2\underline{\hat{a}}_{-\Omega}+2\left(\underline{a}_0\right)^2\underline{\hat{a}}^*_{+\Omega}+\mathrm{i}2\Omega\tau'_{\mathrm{eff}}\left(\left|\underline{a}_0\right|^2\underline{\hat{a}}_{-\Omega}+\left(\underline{a}_0\right)^2\underline{\hat{a}}^*_{+\Omega}\right)\right]. \quad \text{(A11)}$$

We subtract $\mathrm{i}\Delta\phi\underline{\hat{a}}_{+\Omega}$ in Eq. (A10), add $\mathrm{i}\Delta\phi\underline{\hat{a}}_{-\Omega}$ in Eq. (A11) and introduce $\underline{\lambda}' = \lambda+\mathrm{i}\lambda'_i = \lambda+\mathrm{i}\left(\lambda_i-\Delta\phi\right)$. Then, we compute the complex conjugate of the resulting Eq. (A11). We obtain a matrix equation of the form

$$\underline{\lambda}'\begin{pmatrix}\underline{\hat{a}}_{+\Omega}\\ \underline{\hat{a}}^*_{-\Omega}\end{pmatrix} = \begin{pmatrix}-1-\mathrm{i}\zeta-\mathrm{i}d\,\Omega^2+\mathrm{i}\overline{\phi}+\left(\mathrm{i}-r\right)2\left|\underline{a}_0\right|^2 & \left(\mathrm{i}-r\right)\left(\underline{a}_0\right)^2\\ \left(-\mathrm{i}-r\right)\left(\underline{a}^*_0\right)^2 & -1+\mathrm{i}\zeta+\mathrm{i}d\,\Omega^2-\mathrm{i}\overline{\phi}+\left(-\mathrm{i}-r\right)2\left|\underline{a}_0\right|^2\end{pmatrix}\begin{pmatrix}\underline{\hat{a}}_{+\Omega}\\ \underline{\hat{a}}^*_{-\Omega}\end{pmatrix}$$
$$-\frac{\sigma' r\,\tau'_{\mathrm{eff}}\left|\underline{a}_0\right|^2}{\left(\Omega\tau'_{\mathrm{eff}}\right)^2+1}\begin{pmatrix}\left(1+\mathrm{i}\mu\right)\left(\left(\Omega\tau'_{\mathrm{eff}}\right)^2+3\right)\left|\underline{a}_0\right|^2 & \left(1+\mathrm{i}\mu\right)2\left(\underline{a}_0\right)^2\\ \left(1-\mathrm{i}\mu\right)2\left(\underline{a}^*_0\right)^2 & \left(1-\mathrm{i}\mu\right)\left(\left(\Omega\tau'_{\mathrm{eff}}\right)^2+3\right)\left|\underline{a}_0\right|^2\end{pmatrix}\begin{pmatrix}\underline{\hat{a}}_{+\Omega}\\ \underline{\hat{a}}^*_{-\Omega}\end{pmatrix} \quad \text{(A12)}$$
$$+\mathrm{i}\frac{\sigma' r\tau'_{\mathrm{eff}}\left|\underline{a}_0\right|^2}{\left(\Omega\tau'_{\mathrm{eff}}\right)^2+1}2\Omega\tau'_{\mathrm{eff}}\begin{pmatrix}\left(1+\mathrm{i}\mu\right)\left|\underline{a}_0\right|^2 & \left(1+\mathrm{i}\mu\right)\left(\underline{a}_0\right)^2\\ \left(1-\mathrm{i}\mu\right)\left(\underline{a}^*_0\right)^2 & \left(1-\mathrm{i}\mu\right)\left|\underline{a}_0\right|^2\end{pmatrix}\begin{pmatrix}\underline{\hat{a}}_{+\Omega}\\ \underline{\hat{a}}^*_{-\Omega}\end{pmatrix}$$

This equation can be represented as

$$\underline{\lambda}'\begin{pmatrix}\underline{\hat{a}}_{+\Omega}\\ \underline{\hat{a}}^*_{-\Omega}\end{pmatrix} = \begin{pmatrix}m_{A,1}+\mathrm{i}m_{B,1} & m_{A,2}+\mathrm{i}m_{B,2}\\ m^*_{A,2}+\mathrm{i}m^*_{B,2} & m^*_{A,1}+\mathrm{i}m^*_{B,1}\end{pmatrix}\begin{pmatrix}\underline{\hat{a}}_{+\Omega}\\ \underline{\hat{a}}^*_{-\Omega}\end{pmatrix} = \mathbf{M}\begin{pmatrix}\underline{\hat{a}}_{+\Omega}\\ \underline{\hat{a}}^*_{-\Omega}\end{pmatrix}, \qquad \text{where}$$
$$m_{A,1} = -1-\mathrm{i}\zeta-\mathrm{i}d\,\Omega^2+\mathrm{i}\overline{\phi}+\left(\mathrm{i}-r\right)2\left|\underline{a}_0\right|^2-\frac{\sigma' r\,\tau'_{\mathrm{eff}}\left|\underline{a}_0\right|^2}{\left(\Omega\tau'_{\mathrm{eff}}\right)^2+1}\left(1+\mathrm{i}\mu\right)\left(\left(\Omega\tau'_{\mathrm{eff}}\right)^2+3\right)\left|\underline{a}_0\right|^2,$$
$$m_{A,2} = \left(\mathrm{i}-r\right)\left(\underline{a}_0\right)^2-\frac{\sigma' r\,\tau'_{\mathrm{eff}}\left|\underline{a}_0\right|^2}{\left(\Omega\tau'_{\mathrm{eff}}\right)^2+1}\left(1+\mathrm{i}\mu\right)2\left(\underline{a}_0\right)^2, \quad \text{(A13)}$$
$$m_{B,1} = \frac{\sigma' r\,\tau'_{\mathrm{eff}}\left|\underline{a}_0\right|^2}{\left(\Omega\tau'_{\mathrm{eff}}\right)^2+1}2\Omega\tau'_{\mathrm{eff}}\left(1+\mathrm{i}\mu\right)\left|\underline{a}_0\right|^2,$$
$$m_{B,2} = \frac{\sigma' r\,\tau'_{\mathrm{eff}}\left|\underline{a}_0\right|^2}{\left(\Omega\tau'_{\mathrm{eff}}\right)^2+1}2\Omega\tau'_{\mathrm{eff}}\left(1+\mathrm{i}\mu\right)\left(\underline{a}_0\right)^2.$$

Computing the eigenvalues of $\mathbf{M}$ results in

$$\underline{\lambda}_\pm = \Re\left\{m_{A,1}\right\}+\mathrm{i}\Re\left\{m_{B,1}\right\}\pm\underline{\Delta},$$
$$\underline{\Delta} = \left[\left(\Re\left\{m_{A,2}\right\}+\mathrm{i}\Re\left\{m_{B,2}\right\}\right)^2+\left(\Im\left\{m_{A,2}\right\}+\mathrm{i}\Im\left\{m_{B,2}\right\}\right)^2-\left(\Im\left\{m_{A,1}\right\}+\mathrm{i}\Im\left\{m_{B,1}\right\}\right)^2\right]^{1/2}. \quad \text{(A14)}$$

We back-substitute $m_{A,1}, m_{A,2}, m_{B,1}$, $m_{B,2}$, $\lambda_i'$, $\bar{\phi}$, and $\Delta\phi$, and introduce the power of the pumped mode $A = |a_0|^2$ to obtain expressions for the real part and the imaginary port of the gain parameter, see Eq. (9) in the main manuscript. From Eq. (A7) we derive the relation between the pump power $F$ and the power of the pumped mode $A$, see Eq. (10) in the main manuscript.

## B. APPROXIMATION OF THE GAIN PARAMETER FOR TECHNICALLY RELEVANT VALUES FOR TPA, FCA, AND PUMP PARAMETERS

In Section III of the main manuscript we simplify the expression for the real part of the gain parameter $\lambda$ given in Eq. (9) of the main manuscript by assuming technically relevant values for the TPA, FCA, and pump parameters $\{r \approx 1, \sigma' \approx 0.01, \tau_{\text{eff}}' \approx 2\pi \ldots 20\pi, \mu \approx 10, F \approx 10\}$ and $\Omega \approx 10$. Here we specify the magnitude of specific terms occurring in Eq. (9) of the main manuscript for the given parameter ranges. We consider two different cases $\tau_{\text{eff}}' \approx 2\pi$ (left value below the respective term) and $\tau_{\text{eff}}' \approx 20\pi$ (right value below the respective term), and we neglect all terms that are either at least three orders of magnitude smaller than competing terms or that are at least two orders of magnitude smaller and that are approximately constant in $\tau_{\text{eff}}'$. The remaining terms are highlighted in blue. In the following relations, the symbol "$\approx$" used is to be understood as an order-of-magnitude quantification rather than as an approximate equality.

$$\lambda(\Omega) = -\underbrace{1}_{=1} - \underbrace{2rA}_{\approx 2} - \underbrace{r\tau_{\text{eff}}'\sigma' A^2}_{\approx 6\times10^{-2}\ \ldots\ 6\times10^{-1}} \underbrace{\frac{3+(\Omega\tau_{\text{eff}}')^2}{1+(\Omega\tau_{\text{eff}}')^2}}_{\approx 1} + \Re\{\underline{\Delta}\},$$

$$\underline{\Delta} = \left[ \underbrace{A^2}_{\approx 1} \left( \underbrace{r}_{\approx 1} + \underbrace{\frac{2r\tau_{\text{eff}}'\sigma'}{1+(\Omega\tau_{\text{eff}}')^2}}_{\approx 3\times10^{-5}\ \ldots\ 3\times10^{-6}} A - \mathrm{i} \underbrace{\frac{2r\tau_{\text{eff}}'\sigma'\Omega\tau_{\text{eff}}'}{1+(\Omega\tau_{\text{eff}}')^2}}_{\approx 2\times10^{-3}\ \ldots\ 2\times10^{-3}} A \right)^2 \right.$$

$$+ \underbrace{A^2}_{\approx 1} \left( \underbrace{1}_{=1} - \underbrace{\frac{2r\tau_{\text{eff}}'\sigma'\mu}{1+(\Omega\tau_{\text{eff}}')^2}}_{\approx 3\times10^{-4}\ \ldots\ 3\times10^{-5}} A + \mathrm{i} \underbrace{\frac{2r\tau_{\text{eff}}'\sigma'\Omega\tau_{\text{eff}}'\mu}{1+(\Omega\tau_{\text{eff}}')^2}}_{\approx 2\times10^{-2}\ \ldots\ 2\times10^{-2}} A \right)^2$$

$$\left. - \left( \underbrace{\zeta + d\Omega^2 - \frac{\phi_{+\Omega}+\phi_{-\Omega}}{2}}_{\text{Arbitrarily large}} - \underbrace{2A}_{\approx 2} + \underbrace{r\tau_{\text{eff}}'\sigma'\mu A^2}_{\approx 6\times10^{-2}\ \ldots\ 6\times10^{-1}} \underbrace{\frac{3+(\Omega\tau_{\text{eff}}')^2}{1+(\Omega\tau_{\text{eff}}')^2}}_{\approx 1} - \mathrm{i} \underbrace{\frac{2r\tau_{\text{eff}}'\sigma'\Omega\tau_{\text{eff}}'\mu}{1+(\Omega\tau_{\text{eff}}')^2}}_{\approx 2\times10^{-2}\ \ldots\ 2\times10^{-2}} A^2 \right)^2 \right]^{1/2}. \tag{B1}$$

These approximations are equivalent to approximating the optical field $\underline{a}(t',\tau')$ generating free carriers by $|\underline{a}(t',\tau')|^4 \approx |\underline{a}_0|^4$ in Eq (A4), such that only the first term on the r.h.s. of Eq. (A6) remains. The expression for the real part $\lambda$ of the complex gain parameter then reads

$$\lambda(\Omega) \approx -1 - 2rA - r\tau_{\text{eff}}'\sigma' A^2 + \Re\left\{ \sqrt{A^2 r^2 + A^2 - \left( \zeta + d\Omega^2 - \frac{\phi_{+\Omega}+\phi_{-\Omega}}{2} - 2A + r\tau_{\text{eff}}'\sigma'\mu A^2 \right)^2} \right\}. \tag{B2}$$

In absence of AMC, $\phi_{+\Omega} = \phi_{-\Omega} = 0$, this leads to Eq. (11) of the main manuscript.

## C. THRESHOLD PUMP POWER FOR MODULATION INSTABILITY: APPROXIMATE ANALYTICAL VS. NUMERICAL EVALUATION

In Figure 2 of the main manuscript, we analytically evaluate an approximation of the threshold pump power $F_{\mathrm{th}}$ need to achieve modulation instability, based on realistic values for $\tau'_{\mathrm{eff}}, A, \Omega, r, \sigma', \mu$, and for anomalous group-velocity dispersion $d > 0$ with $d\,\Omega^2_{\max} - A = 0$ at $\lambda(\Omega_{\max}) = 0$. We further assume that mode coupling is absent or sufficiently weak such that avoided mode crossings do not need to be considered ($\phi_{\Omega'} = 0 \;\forall\; \Omega'$). In the following, we validate this approximation by a numerical investigation. To this end, we find the threshold pump power $F_{\mathrm{th}}$ by minimizing $F$ according to Eq. (10) of the main manuscript when varying $A$, $\Omega$ and $\zeta$ under the constraint $\lambda(\Omega_{\max}) = 0$, Eq. (9). We choose $d = 0.0025$, $\mu = 7.5$ and sweep the quantities $r$ and $\tau'_{\mathrm{eff}}\sigma'$ to cover the same parameter space as in Fig. 2. Figure C1(a) shows the result of the analytic approximation and corresponds to Fig. 2 of the main manuscript, while Fig. C1(b) displays the numerical results. No difference is visible from the two graphs. The relative difference $|\Delta F_{\mathrm{th}}|/F_{\mathrm{th,full}} = |F_{\mathrm{th,simple}} - F_{\mathrm{th,full}}|/F_{\mathrm{th,full}}$ is smaller than $10^{-2}$ in the whole region, and smaller than $10^{-3}$ for $r \le 0.4$, see Figure C1(c).

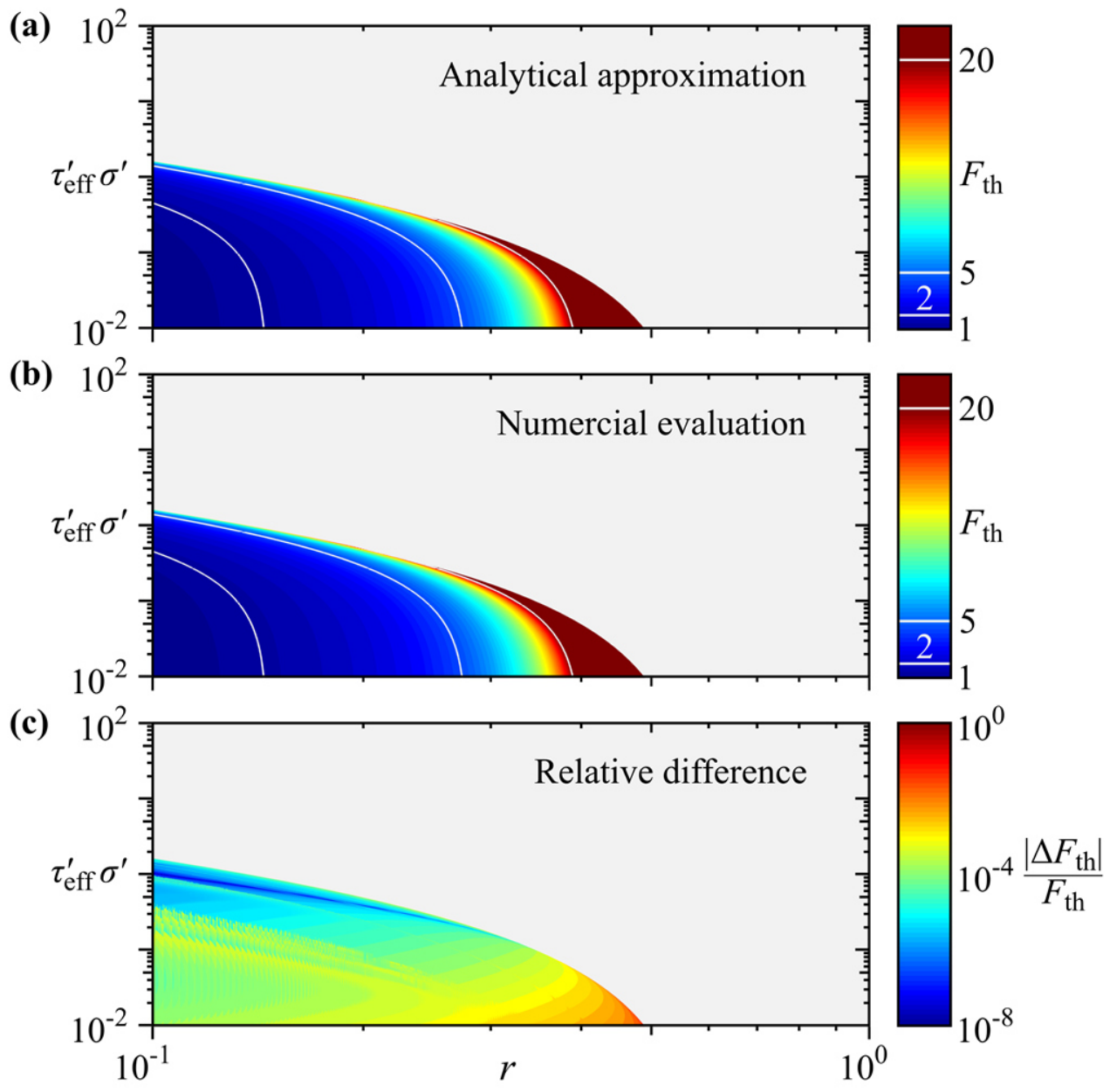

**Figure C1:** Threshold pump power needed to achieve modulation instability. (a) Approximate analytical evaluation, see also Fig. 2 of the main manuscript. (b) Corresponding numerical calculation. (c) Relative color-coded difference in threshold pump powers when comparing the analytical and the numerical method. The deviation is smaller than 1% in the whole region, and smaller than $10^{-3}$ for a TPA parameter $r \le 0.4$.

## D. MODE FIELD SIMULATIONS FOR NONLINEARITY PARAMETER, GROUP REFRACTIVE INDEX AND DISPERSION PARAMETER OF SILICON-PHOTONIC WAVEGUIDES

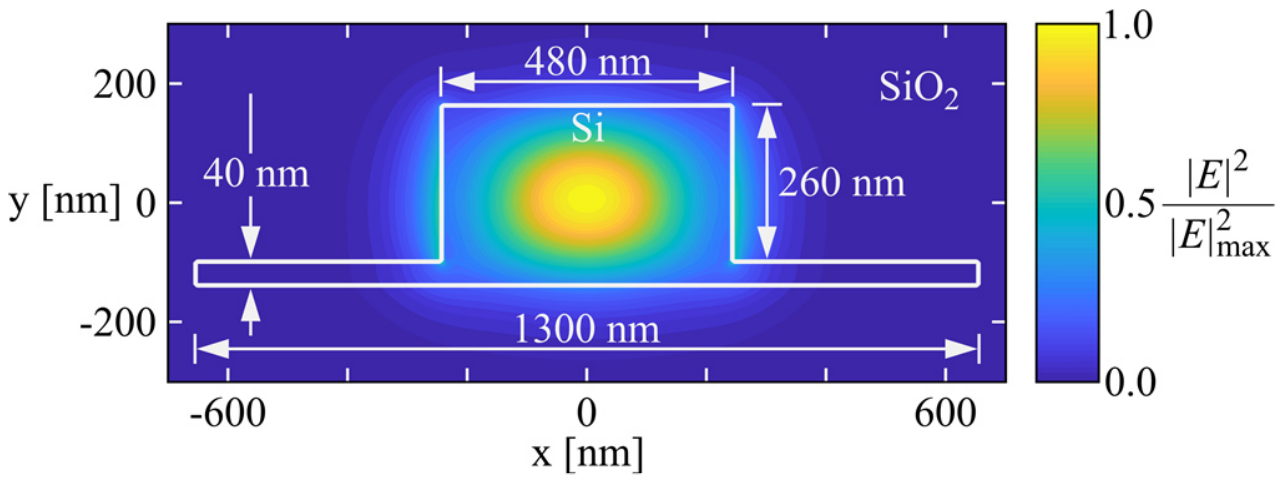

**Figure D1:** Normalized modulus squared of the total electric field of the quasi-TE mode at 1550 nm. We assume a waveguide bent to the left with bend radius of 115 µm. The white outline marks the undoped silicon part of the waveguide.

To determine the waveguide nonlinearity and the dispersion of the waveguide presented in Fig. 1 of the main manuscript at a pump wavelength of 1550 nm, we perform finite-element simulations using commercial tools (RSoft Photonics CAD Suite, FemSIM). The waveguide in these simulations is either straight or has a bend radius of 115 µm, corresponding to the resonator design shown in Fig. 1 of the main manuscript. The refractive indices of silica $n_{SiO_2}$ and silicon $n_{Si}$ are included according to Sellmeier's equation. The modulus squared of the electric field of the quasi-TE mode field within the silicon waveguide is shown in Fig. D1. The undoped silicon portion of the waveguide is marked with a white outline, and the mode is essentially confined to this region such that the optical field will not be affected by the *p*-*i*-*n*-junction and by the vertical interconnect accesses (vias) shown in Fig. 1 of the main manuscript. Since the bend radius $R_B >> w$ is large relative to the waveguide width $w$ and because $\sqrt{n_{Si}^2 - n_{SiO_2}^2}\, R_B >> \lambda$ holds, the field appears symmetrical in the horizontal direction.

Given the field distributions $\mathbf{E}(x,y), \mathbf{H}(x,y)$, we compute the nonlinear waveguide parameter $\gamma$ using [48]

$$\gamma = \frac{\omega_P \varepsilon_0 n_{Si}^2 n_{2,Si}}{Z_0} \frac{\iint_{Si} \left|\mathbf{E}(x,y)\right|^4 \mathrm{d}x\,\mathrm{d}y}{\left|\iint_{total} \mathrm{Re}\left\{\mathbf{E}(x,y) \times \mathbf{H}^*(x,y)\right\} \cdot \mathbf{e}_z \,\mathrm{d}x\,\mathrm{d}y\right|^2}. \tag{D1}$$

Here, $\omega_P$ is the angular frequency of the optical pump field, $Z_0$ is the free-space wave impedance, and $\mathbf{e}_z$ the unit vector along a perimeter of the ring-shaped waveguide. We assume that the third-order nonlinear susceptibility of silicon is a scalar and can be expressed by corresponding Kerr coefficient $n_{2,Si} = 6.5 \times 10^{-18}\,\mathrm{m^2\,W^{-1}}$ [21]. The resulting nonlinearity parameter at a wavelength of 1550 nm amounts to $\gamma = 257\,\mathrm{W^{-1}\,m^{-1}}$. In order to determine the dispersion coefficient, mode field simulations are performed for varying wavelengths in a spectral range of 60 nm (7.5 THz) around 1550 nm. From these simulations, we obtain the effective refractive index, which is then used to determine the group refractive index $n_g = 4.15$ and the dispersion coefficient $\beta_2 = -587\,\mathrm{ps^2\,km^{-1}}$ both at a wavelength of 1550 nm.

## E. INTEGRATION OF THE LUGIATO-LEFEVER EQUATION AND THE FREE-CARRIER EQUATION

In Section V of the main manuscript, we numerically integrate the normalized Equations (5) and (6) of the main manuscript using the split-step Fourier method [49]. The normalized temporal time step is $\Delta t'$, the dispersion operator is denoted by $\hat{D}$, and the nonlinear operator is $\hat{N}$:

$$\begin{aligned} \underline{a}(t'+\Delta t',\tau') &= \sqrt{F}\Delta t' + \mathrm{e}^{\hat{N}(t',\tau')\Delta t'/2}\,\mathrm{e}^{\hat{D}\Delta t'}\,\mathrm{e}^{\hat{N}(t',\tau')\Delta t'/2}\,\underline{a}(t',\tau'), \\ \hat{D} = -1-\mathrm{i}\zeta-\mathrm{i}d\frac{\partial^2}{\partial\tau'^2}, &\quad \hat{N}(t',\tau') = (\mathrm{i}-r)\left|\underline{a}(t',\tau')\right|^2 - \sigma'(1+\mathrm{i}\mu)N_\mathrm{c}'(t',\tau'). \end{aligned} \tag{E1}$$

The linear differential equation Eq. (6) in the main manuscript for the free-carrier density $N_\mathrm{c}'(t',\tau')$ is solved by using a frequency-discrete Fourier transform (Fourier series):

$$\begin{aligned} N_\mathrm{c}'(t',\tau') &= r\tau'_\mathrm{eff}\mathcal{F}^{-1}\left\{\frac{\mathcal{F}\left\{\left|\underline{a}(t',\tau')\right|^4\right\}}{1+\mathrm{i}u\,\tau'_\mathrm{eff}}\right\}, \\ \mathcal{F}\{y(\tau')\}_u = \breve{y}_u = \frac{1}{2\pi}\int_0^{2\pi} y(\tau')\mathrm{e}^{-\mathrm{i}u\tau'}\,\mathrm{d}\tau', &\quad y(\tau') = \mathcal{F}^{-1}\{\breve{y}\} = \sum\nolimits_u \breve{y}_u\,\mathrm{e}^{\mathrm{i}u\tau'}. \end{aligned} \tag{E2}$$

In these relations, the discrete frequency index $u = 0, \pm1, \pm2, \pm3, \ldots$ accounts for the various comb tones, which superimpose to a waveform $y(\tau')$ which is periodic in fast time $\tau' \in [0, 2\pi)$. We substitute Eq. (E2) in Eq. (E1) and integrate stepwise.

## F. DISPERSION PROFILE OF COUPLED MODE FAMILIES

In this section we describe the derivation of the dispersion profile used in Section V of the main manuscript to simulate comb formation in a normal-dispersive microresonator under the influence of an avoided mode crossing, see Column (d) of Fig. 3. In a first step, we consider two unperturbed mode families M1 and M2. The comb is generated in the resonances of mode family M1, while mode family M2 may induce local shifts of resonance positions of M1. The mode families are described by the frequency of the central mode $\omega_{0,\mathrm{M1}}$, $\omega_{0,\mathrm{M2}}$ and by the corresponding free spectral ranges $f_{\mathrm{FSR,M1}}$, $f_{\mathrm{FSR,M2}}$, such that the respective resonance frequencies read

$$\begin{aligned} \omega_{\Omega',\mathrm{M1}} &= \omega_{0,\mathrm{M1}} + \Omega'\times(2\pi\times f_{\mathrm{FSR,M1}}) - \beta_2 L/(2t_\mathrm{R})\left(\Omega'\times(2\pi\times f_{\mathrm{FSR,M1}})\right)^2, \\ \omega_{\Omega',\mathrm{M2}} &= \omega_{0,\mathrm{M2}} + \Omega'\times(2\pi\times f_{\mathrm{FSR,M2}}). \end{aligned} \tag{F1}$$

Here, we have omitted dispersive terms of the second mode family for simplicity. The dispersion profiles of the modes describe the deviation of the various resonance frequencies from an equidistant grid defined by the center free spectral range $f_{\mathrm{FSR,M1}}$ of mode family M1. Using Eq. (F1), the dispersion profiles can be stated for the two mode families M1 and M2,

$$\begin{aligned} \omega_{\Omega',\mathrm{M1}} - \omega_{0,\mathrm{M1}} - \Omega'\times(2\pi\times f_{\mathrm{FSR,M1}}) &= -\beta_2 L/(2t_\mathrm{R})\left(\Omega'\times(2\pi\times f_{\mathrm{FSR,M1}})\right)^2, \\ \omega_{\Omega',\mathrm{M2}} - \omega_{0,\mathrm{M1}} - \Omega'\times(2\pi\times f_{\mathrm{FSR,M1}}) &= \omega_{0,\mathrm{M2}} - \omega_{0,\mathrm{M1}} - \Omega'\times 2\pi\left(f_{\mathrm{FSR,M2}} - f_{\mathrm{FSR,M1}}\right). \end{aligned} \tag{F2}$$

The dispersion profile of M1 is a set of discrete points located on a parabola, which is defined by the second order dispersion coefficient of mode family M1. It allows to compute the phase deviation $\varphi(\Omega')$ accumulated by comb modes $\Omega'$, see Eq. (15) in the main text. In contrast to the parabolic dispersion profile of mode family M1, the points given by the dispersion profile of M2 are located on a strongly inclined line, which is indicated in red in Fig. 3(d), Row R2 of the main manuscript. Unavoidable deviations from the ideal resonator geometry lead to coupling of the mode families leads and hence to a hybridization, which is accompanied by a local shift of the resonance frequencies from $\omega_{\Omega',\mathrm{M1/M2}}$ to two hybrid mode resonances $\omega_{\Omega',\pm}$ for each mode index $\Omega'$ [43]. This shift depends on the coupling strength $\theta$ between the respective modes , such that the hybrid mode resonance frequencies are given by [43]

$$\omega_{\Omega',\pm} = \frac{\omega_{\Omega',\mathrm{M1}} + \omega_{\Omega',\mathrm{M2}}}{2} \pm \sqrt{\theta^2 + \frac{\left(\omega_{\Omega',\mathrm{M1}} - \omega_{\Omega',\mathrm{M2}}\right)^2}{4}}. \tag{F3}$$

For these hybrid modes, the dispersion profile is given in the same manner as for $\omega_{\Omega',\mathrm{M1/M2}}$ by $\omega_{\Omega',\pm} - \omega_{0,\mathrm{M1}} - \Omega' \times \left(2\pi \times f_{\mathrm{FSR,M1}}\right)$. Mode family M1 is the relevant mode family for comb formation in our system. For the gain parameter computation and the simulation of comb formation, we need to choose a certain set of resonances that are close to those of mode family M1 and that contain either the up-shifted frequencies $\omega_{\Omega',+}$ or the down-shifted frequencies $\omega_{\Omega',-}$ for every mode index $\Omega'$. For large differences $\omega_{\Omega',\mathrm{M1}} - \omega_{\Omega',\mathrm{M2}}$ of resonance frequencies of the native modes M1 and M2, the native resonance $\omega_{\Omega',\mathrm{M1}}$ corresponds to the hybrid resonance $\omega_{\Omega',+}$ in case $\omega_{\Omega',\mathrm{M1}} - \omega_{\Omega',\mathrm{M2}} > 0$ and to the hybrid resonance $\omega_{\Omega',-}$ if $\omega_{\Omega',\mathrm{M1}} - \omega_{\Omega',\mathrm{M2}} < 0$. Near the mode coupling, we choose either $\omega_{\Omega',+}$ or $\omega_{\Omega',-}$ such that the obtained dispersion profile resembles a typical avoided mode crossing that exhibits approximately symmetric deviations from the native dispersion profile of mode family M1, as observed in experiments, see, e.g., [43]. From the selected hybrid resonances, we determine the resonance shifts $\delta\omega_{\Omega',+} = \omega_{\Omega',\mathrm{M1}} - \omega_{\Omega',+}$ and $\delta\omega_{\Omega',-} = \omega_{\Omega',\mathrm{M1}} - \omega_{\Omega',-}$ respectively. These shifts are used for numeric simulations, the computation of the gain parameter according to Eq. (9) of the main text and the determination of phase deviations $\varphi'(\Omega')$ according to Eq. (15). Generally, the crucial parameter for modulation instability is the maximum resonance frequency shift $\omega_{\Omega',\mathrm{M1}} - \omega_{\Omega',\pm}$, which amounts to $-2\pi \times 630\,\mathrm{MHz}$ or approximately $1\%$ of the FSR of the microresonator under investigation. This is in accordance with experimentally observed resonance shifts [43].

## G. LIST OF PHYSICAL QUANTITIES. PARAMETERS FOR THE SIMULATION OF FREQUENCY COMB DYNAMICS

Tables G1 summarizes both the physical and the normalized quantities used for describing the dynamics of the optical field in the Kerr-nonlinear microresonators. In Table G2, we specify physical and normalized microresonator parameters along with their numerical values and the associated source references.

**Table G1:** Physical and normalized quantities.

| Physical quantity | Symbol | Value | Source | Symbol of equivalent normalized quantity |
|---|---|---|---|---|
| Angular pump frequency | $\omega$ | $2\pi \times 193.41\,\text{THz}$ | Assumption | - |
| Pump power | $P_{\text{in}}$ | $12 \ldots 50\,\text{mW}$ | Assumption | $F$ |
| Detuning | $(\omega_{\text{R}} - \omega_{\text{P}})t_{\text{R}}$ | $(1.2\ldots2)\,\text{GHz} \times 2\pi\, t_{\text{R}}$ | Assumption | $\zeta$ |
| Slow time (natural time) | $t$ | – | – | $t'$ |
| Fast time (retarded time) | $\tau$ | – | – | $\tau'$ |
| Optical field | $\underline{E}$ | – | – | $\underline{a}$ |
| Free carrier density | $N_{\text{c}}$ | – | – | $N'_{\text{c}}$ |
| Sideband frequencies | $\omega_{\Omega'}$ | – | – | $\Omega'$ |
| Gain rate | – | – | – | $\underline{\lambda}$ |
| Phase deviations | $\varphi(\Omega')$ | | | $\varphi'(\Omega')$ |

**Table G2:** Microresonator parameters and normalized quantities.

| Microresonator parameter | Symbol | Value | Source | Symbol of equivalent normalized quantity | Further occurrence in normalized quantities |
|---|---|---|---|---|---|
| Power loss per length | $\alpha_i$ | $46\,\mathrm{m}^{-1}\left(2\,\mathrm{dB\,cm}^{-1}\right)$ | [21] | – | $d,\sigma',F,\zeta,t',\underline{a},N_c',\phi_{\Omega'}$ |
| Group refractive index | $n_\mathrm{g}$ | 4.15 | Simulation | – | $F,\zeta,t',\phi_{\Omega'}$ |
| Second-order dispersion coefficient | $\beta_2$ | $-587\,\mathrm{ps}^2\,\mathrm{m}^{-1}$ | Simulation | $d$ | – |
| Nonlinearity parameter | $\gamma$ | $257\,\mathrm{W}^{-1}\,\mathrm{m}^{-1}$ | Simulation | – | $F,a$ |
| Kerr coefficient | $n_2$ | $6.5\times10^{-18}\,\mathrm{m}^2\,\mathrm{W}^{-1}$ | [21] | – | $r,\sigma',N_c'$ |
| TPA parameter | $\beta_\mathrm{TPA}$ | $0.7\times10^{-11}\,\mathrm{m\,W}^{-1}$ | [21] | $r$ | – |
| FC dwell time | $\tau_\mathrm{eff}$ | $12\ldots100\,\mathrm{ps}$ | [21] | $\tau'_\mathrm{eff}$ | – |
| FC cross-section | $\sigma$ | $1.45\times10^{-21}\,\mathrm{m}^2$ | [21] | $\sigma'$ | – |
| FC dispersion parameter | $\mu$ | 7.5 | [17] | – | – |
| Confinement factor | $\Gamma_\mathrm{c}$ | 1 | Assumption | – | – |
| Round-trip time | $t_\mathrm{RT}$ | $10\,\mathrm{ps}$ | Assumption | – | $d,\sigma',F,\zeta,t',\underline{a},N_c'$ |
| Free spectral range | $f_\mathrm{FSR}=t_\mathrm{RT}^{-1}$ | $100\,\mathrm{GHz}$ | Assumption | – | See round-trip time |
| Local resonance shift | $\delta\omega_{\Omega'}$ | Max. $2\pi\times630\,\mathrm{MHz}$ | Assumption | $\phi_{\Omega'}$ | – |